\documentclass[aps,preprint,prd,nofootinbib]{revtex4-1}
\usepackage{natbib}
\usepackage{times}
\usepackage{amssymb,amsbsy,amsmath,amsfonts}
\usepackage{graphicx}

\usepackage{float}

\usepackage{rotating}
\usepackage{ulem} 
\usepackage{slashed}
\usepackage{morefloats}
\usepackage{rotating}
\usepackage{srcltx}
\usepackage{subfigure}
\usepackage{color}

\begin{document}

\newcommand{\eq}[1]{{(\ref{#1})}}
\preprint{UCSD/PTH 14-09}

\title{
Reassessing the discovery potential of the $B \to K^{*} \ell^+\ell^-$ 
decays in the large-recoil region: SM challenges and BSM opportunities}

\author{S. J\"ager$^{1}$}
\author{J. Martin Camalich$^{2,3}$}

\affiliation{
$1$Department of Physics and Astronomy, University of Sussex;
Brighton BN1 9QH, United Kingdom\\
$^2$Dept. Physics, University of California, San Diego, 9500 Gilman Drive, 
La Jolla, CA 92093-0319, USA\\
$^3$PRISMA Cluster of Excellence Institut f\"ur Kernphysik, 
Johannes Gutenberg-Universit\"at Mainz, 55128 Mainz, Germany}

\begin{abstract}
We critically examine the potential to disentangle Standard
Model (SM) and New Physics (NP) in $B \to K^* \mu^+\mu^-$
and $B\to K^* e^+ e^-$ decays, focusing on $(i)$ the LHCb anomaly,
$(ii)$ the search for right-handed
currents, and $(iii)$ lepton-universality violation. Restricting 
ourselves to the large-recoil region, we advocate a parameterisation of
the hadronic matrix elements that separates model-independent
information about nonperturbative QCD from the results 
of model calculations. We clarify how to estimate corrections to
the heavy-quark limit that would generate a right-handed (virtual) photon
in the $b\to s\gamma$ contribution to the decay. We then apply this
approach to the discussion of various sets of observables of
increasing theoretical cleanness. First, we show that angular observables
in the optimized $P_i^{(\prime)}$ basis are, in general, still not robust
against the long-distance QCD effects, both numerically and by examining analytically the
dependence on corrections to the (model-independent) heavy-quark limit.
As a result, while a fit to data favours a NP contribution to the semileptonic operators 
of the type $\delta C_9\simeq-1.5$, this comes at a relatively small statistical significance of 
$\lesssim2 \sigma$, once such power corrections are properly accounted for.
Second, two of these observables, $P_1$ and $P_3^{CP}$ are particularly
clean at very low $q^2$ and sensitive probes of right-handed
quark currents. We discuss their potential to set stringent bounds
on the Wilson coefficient $C_7^\prime$, especially using data of the electronic
mode, and we update the bounds with current angular data in 
the muonic channel. Finally, in light of the recent hint of
lepton-universality violation in $B^+\to K^+\ell\ell$,
we introduce and investigate new lepton-universality observables
involving angular observables of the muonic and electronic modes and
their zero crossings, and show that, if the effect is of the size suggested by
experiment, these can clearly distinguish between different NP explanations
in terms of underlying semileptonic operators.
\end{abstract}

\maketitle

\section{Introduction}

Rare $B$ decays such as $B \to M \ell^+ \ell^-$, where $M$ is a
charmless hadronic state, are powerful probes of
Beyond-Standard Model (BSM) physics due to their short-distance
sensitivity combined with their GIM and CKM suppression in the
Standard Model (SM). In case $M$ is a vector resonance, like the
$K^*(892)$ ($K^*$ from now on), the decays have a very rich
kinematical structure that leads to up to twenty-four angular
observables  (including direct $CP$ asymmetries) which are
functions of the dilepton invariant mass squared, $q^2$~\cite{Ali:1991is,
Burdman:1998mk,Melikhov:1998cd,Kruger:1999xa,Beneke:2001at,Bosch:2001gv,
Kagan:2001zk,Feldmann:2002iw,Beneke:2004dp,Grinstein:2004vb,Kruger:2005ep,
Bobeth:2008ij,Egede:2008uy, Altmannshofer:2008dz,Bobeth:2010wg, 
Egede:2010zc,Alok:2010zd,Beylich:2011aq,Bobeth:2011gi,Becirevic:2011bp,
Alok:2011gv,Lu:2011jm,Matias:2012xw,Becirevic:2012dp,Korchin:2012kz,
Das:2012kz,Matias:2012qz,Blake:2012mb,DescotesGenon:2011yn,
Beaujean:2012uj,DescotesGenon:2012zf,Bobeth:2012vn,Jager:2012uw,
Descotes-Genon:2013vna,Doring:2013wka,Horgan:2013pva,
Lyon:2014hpa,Descotes-Genon:2014uoa}.

Experimental results on $ B \to K^* \ell^+ \ell^-$
include measurements of the branching fraction, forward-backward 
asymmetry and the longitudinal polarization fractions by the 
$B$-factories~\cite{Aubert:2008bi,Aubert:2008ps,Wei:2009zv,Lees:2012tva},
CDF~\cite{Aaltonen:2011cn}, LHCb~\cite{Aaij:2013hha,Aaij:2013iag}, 
CMS~\cite{Chatrchyan:2013cda} and ATLAS~\cite{Schieck:2013tsa}; 
measurements including the angular observables $A_T^{(2)}$ and $A_{\rm im}$ have 
been done by CDF \cite{Aaltonen:2011ja} and LHCb~\cite{Aaij:2013iag}.
Intriguingly, LHCb reported a $3.7\sigma$ discrepancy
with the SM in the muonic mode ($\ell=\mu$) in the course of the first complete
($CP$-averaged) angular analysis of the final state system~\cite{Aaij:2013qta}. 
The putative effect occurs in the $[4.3, 8.68]$ GeV$^2$
dilepton mass bin of the angular observable $P_5'$, but also
with a lower significance of $2.5\sigma$ in the bin  
$[1, 6]$ GeV$^2$. Other tensions have been pointed out
in the data ($P_2$~\cite{Descotes-Genon:2013wba} or 
$F_L$~\cite{Altmannshofer:2013foa}) and global 
analyses of $b\to s\mu\mu$ and $b\to s\gamma$ decays have claimed the
data to be in tension with the SM with a statistical significance
of up to $4.5\sigma$~\cite{Descotes-Genon:2013wba}.

Beyond the SM, this tension can be ascribed to a negative shift of the Wilson
coefficient of the semileptonic-vector operator 
$Q_9$~\cite{Descotes-Genon:2013wba} in the weak
Hamiltonian, although contributions to other Wilson coefficients
have also been discussed~\cite{Descotes-Genon:2013wba,Altmannshofer:2013foa,Beaujean:2013soa,
Hurth:2013ssa,Altmannshofer:2014rta}. On the other hand, a previous analysis of the
angular observables using a model-independent parametrization of
the hadronic uncertainties~\cite{Jager:2012uw} that minimized the input from 
nonperturbative calculations, led to SM predictions with theoretical errors considerably larger
than those employed in the above-mentioned fits. More recently, a Bayesian analysis
of the data  found a good agreement with the SM~\cite{Beaujean:2013soa},
allowing the hadronic parameters to float in the fit.

Besides that, the LHCb recently reported an even more surprising 
deficit of $B^+\to K^+\mu\mu$ decays as compared to $B^+\to K^+e\,e$, 
with a significance of $2.6\sigma$~\cite{Aaij:2014ora}. 
This signal of lepton-universality violation (LUV) has been analysed 
by different groups 
~\cite{Alonso:2014csa,Hiller:2014yaa,Ghosh:2014awa,Biswas:2014gga,Hurth:2014vma,Glashow:2014iga,
Hiller:2014ula} with the common conclusion that the only plausible sources of
this effect are the semileptonic operators $Q^{(\prime)}_{9,10}$.
Moreover, it was recently pointed out~\cite{Hiller:2014ula} that similar
deficits in the inclusive $b\to s\ell\ell$ decay have been observed by  
Belle~\cite{Belle} and BaBar~\cite{Lees:2013nxa}.

Because of the far-reaching implications of potential manifestations
of new-physics (NP), a clear understanding of the SM ``background''
expectation is needed. In this work, we critically re-examine 
the anatomy of the uncertainties in the theoretical description of 
$B \to K^* \ell^+ \ell^-$ decays in the large-recoil region and 
classify different observables according to their theoretical cleanness. 
Our presentation draws on our longer work~\cite{Jager:2012uw},
which was focused on a transparent decomposition of the decay amplitudes into
(perturbatively) calculable and nonperturbative ingredients, a general
but minimal parametrization of the latter, and a discussion
of present and prospective knowledge of those parameters.
The framework in \cite{Jager:2012uw} was tailored to the lower endpoint region
of the dilepton invariant mass spectrum. We update it here and, with minor
extensions, we apply to the case at hand.

We then update our SM predictions for the angular distribution of
$B \to K^* \mu^+ \mu^-$ and provide new predictions for the $B \to K^* e^+ e^-$
angular distribution. The complex dependence that observables in the $P_i^{(\prime)}$
basis in general have on the underlying hadronic matrix elements is illustrated
analysing the tension of the $b\to s\mu\mu$ data with the SM. We further discuss 
two observables, $P_1$ and $P_3^{CP}$, which are particularly clean and that
can be used to put stringent bounds on electromagnetic operators induced by
right-handed currents. Finally, in light of the LUV signals hinted by different
$b\to s\ell\ell$ measurements we introduce and study new lepton-universality ratios
which are very accurately predicted in the SM. We show that they provide an excellent
benchmark to confirm and characterize the effect and we study prospects 
in different NP scenarios.

\section{Connecting short-distance physics to observables}
\label{sec:SDtoObs}

The decay $\bar{B} \to \bar{K}^* \ell^+ \ell^-$ 
proceeds via the $\Delta B=1$ weak Hamiltonian (see e.g.
\cite{Buchalla:1995vs}), which encapsulates short-distance SM contributions
from scales above $\mu \sim m_b$,
as well as any NP with mass scale beyond the weak scale,
in a set of Wilson coefficients.
In the SM, the lepton pair is always produced through either the
leptonic vector or axial vector current.
The three axial vector helicity amplitudes are:
\begin{equation} \label{eq:HA}
   H_A(\lambda)  = - i N \tilde V_\lambda(q^2) C_{10} ,
\end{equation}
where $\lambda$ is the helicity of the $\bar{K}^*$ and $N$ a
normalisation constant. They receive contributions only from the
semileptonic part of the weak Hamiltonian and factorize ``naively'' 
into helicity form factors $\tilde{V}_\lambda(q^2)$
(\cite{Bharucha:2010im,Jager:2012uw};
conventions in this paper follow \cite{Jager:2012uw})
and the Wilson coefficient $C_{10}$. 

If the lepton mass is not neglected, the axial-vector current can also create the
dilepton in a pseudoscalar state ($\lambda = 0$ only), bringing in another,
naively factorizing, amplitude
$H_P(q^2) = -i N \frac{2\, m_\ell |\vec{k}|}{q^2}\frac{m_b}{m_b+m_s} S(q^2) C_{10}$,
(equivalent to what is often called ``timelike'' amplitude), and one extra, 
scalar form factor $S$; $|\vec k| = \lambda^{1/2}(m_B^2,
m_{K^*}^2, q^2)/(2 m_B)$ is the momentum of the vector meson in the
$B$-meson rest frame. 

In addition, the dilepton can be produced through the vector leptonic current.
The corresponding three vector helicity amplitudes $H_V(\lambda)$ again receive
contributions from the semileptonic $\Delta B=1$
Hamiltonian. However, they comprise further terms originating in the magnetic penguin
operator $Q_{7\gamma}$, as well as from the hadronic part of the weak
Hamiltonian, whereby the dilepton is created through a virtual photon.
The former bring in a further set of three form factors,
while the latter contributions include ``charm loops'',
``annihilation'', etc, and do not factorize naively.
In the notation of \cite{Jager:2012uw}:
\begin{equation} \label{eq:HV}
  H_V(\lambda) = - i N \Bigg[ \tilde V_\lambda(q^2) C_9
   + \frac{2\,m_b m_B}{q^2} \tilde T_\lambda(q^2) C_7 - \frac{16\pi^2
     m_B^2}{q^2} h_\lambda(q^2) \Bigg].
\end{equation}

Beyond the SM, the helicity amplitudes may receive extra contributions
from modified Wilson coefficients $C_7$, $C_{9}$, $C_{10}$, 
as well as the chirally-flipped operators, $Q_7'$, 
$Q_{9}'$, $Q_{10}'$, if present. Furthermore,
in the most general NP scenario there will a be further ``scalar'' and
three ``tensor'' amplitudes~\footnote{The NP scalar contributions to 
$B\rightarrow K^*\ell^+\ell^-$ are tightly bounded by the pure 
leptonic rare decay $B_s\rightarrow \ell^+\ell^-$~\cite{Altmannshofer:2008dz,
Alonso:2014csa}. Tensor operators can be neglected if we assume the scale of
NP to be well above the electro-weak scale~\cite{Alonso:2014csa}.}. 

We would like to emphasize the simplicity and transparency 
of \eq{eq:HA} and \eq{eq:HV} when compared to the more traditional 
transversity amplitudes involving chiral lepton currents. 
In particular, it exposes in a clear way the various hadronic uncertainties 
impacting on the vector helicity amplitudes, 
which stem from two rather than one form factors (per helicity),
$\tilde V_\lambda$ and $\tilde T_\lambda$, and, in addition, the nonlocal 
correlator $h_\lambda$. While the photon-pole
dominance can be exploited to identify especially clean null tests of the
SM at very low $q^2$, particular care will be needed in attributing BSM
effects to observables that involve the vector helicity amplitudes 
(e.g. through $C_9$) away from the endpoint of the large-recoil
region.

Finally it is worth recalling that the residues of the vector helicity amplitudes
are related to the amplitude of the radiative decay:
\begin{eqnarray}
  {\cal A}(\bar B \to K^*(\lambda) \gamma(\lambda))
    &=& \lim_{q^2 \to 0} \frac{q^2}{e} H_V(q^2=0; \lambda) \nonumber \\
    &=& \frac{i N m_B^2}{e} \left[\frac{2 m_b}{m_B}
          (C_{7} \tilde{T}_\lambda(0) - C_{7}^\prime \tilde{T}_{-\lambda}(0))
          - 16 \pi^2 h_\lambda(q^2=0) \right]   \label{eq:radiative}.
\end{eqnarray}

\subsection{Minimal parametrization of nonperturbative QCD}
\label{sec:theo}
 
The key to understanding the uncertainties on various observables is a
full and transparent description of the nonperturbative input, both of
the factorizable (form factor) and (naively) nonfactorizable ($h_\lambda$)
type. This allows to disentangle model-independent  constraints 
from assumptions in the modelling of nonperturbative QCD effects.

A considerable simplification of the nonperturbative dynamics arises in the
combined heavy-quark and large-energy (HQ/LE) limit
of QCD~\cite{Dugan:1990de,
Charles:1998dr,Beneke:1999br,Beneke:2000ry,Bauer:2000yr,Bauer:2001yt,Beneke:2002ph}, 
corresponding to the large-recoil or low-$q^2$ region of the decay. In this limit, both the form factors
\cite{Charles:1998dr,Beneke:2000wa,Beneke:2003pa} and the
(naively) nonfactorizable
term $h_\lambda$ \cite{Beneke:2001at,Beneke:2004dp} exhibit QCD
factorization (collinear factorization) into universal ``soft'' form
factors, light-cone distribution amplitudes (LCDA), and perturbatively
calculable hard kernels; this structure is most transparent when
formulated in soft-collinear effective field theory (SCET) 
where the hard kernels become Wilson coefficients of operators built
out of collinear and soft fields~\cite{Bauer:2000yr,Bauer:2001yt,Beneke:2002ph}. 
In particular, the number of independent form factors is reduced
from seven to two, the vector and tensor helicity form factors in
\eq{eq:HV} are related, and two helicity amplitudes vanish altogether.

A fundamental limitation to factorization is the fact that ${\cal O}(\Lambda/E, \Lambda/m_B)$ power
corrections (to be denoted generically by ${\cal O}(\Lambda/m_B)$ from now on), 
do not factorize; attempts to do so lead to end-point divergent convolutions.
Therefore, besides the parametric uncertainties entering the amplitude 
in the exact HQ/LE limit, one needs to take into account these
power corrections. In this section we update and extend
the model-independent treatment of power corrections
introduced for the first time in~\cite{Jager:2012uw}. We stress at the
outset that this issue cannot be sidestepped by employing form factor
calculations in the light-cone sum rule (LCSR) framework (or other existing
frameworks). To the extent that
LCSR calculations give an unambiguous, controlled (i.e.\ systematically
improvable) approximation, they involve convolutions of perturbative
kernels with LCDA and a twist expansion
very similar, and underpinned by very similar Feynman diagram
calculations, as those applying to the
heavy-quark expansion. However, these are necessarily accompanied
by model-dependent steps before a hadronic quantity can be extracted,
most importantly a modelling of an infinite tower of continuum
contributions. While there is a standard convention for attaching
uncertainties due to this, the procedure is quantitatively justified only by a
number of numerical successes. We will however
clarify in what sense LCSR calculations can be used to estimate
\textit{corrections} to the HQ/LE limit, for which much less
relative accuracy is required. This was shown in
detail for the helicity $+1$ form factors in~\cite{Jager:2012uw}.
Below we will also clarify further in what sense the leading corrections to the
heavy-quark limit of $h_+$ can be precisely identified with a matrix
element that can be estimated with the LCSR method as done
in~\cite{Jager:2012uw}. 

\subsubsection{Form factors}
\label{sec:theoff}

We start rescaling the helicity-zero form factors:
\begin{equation}
  V_0(q^2) = \frac{\sqrt{q^2}}{|\vec k|} \tilde V_0(q^2), \quad
  T_0(q^2) = \frac{m_B^2}{\sqrt{q^2} |\vec k|} \tilde T_0(q^2), \quad
  V_\pm(q^2) = \tilde V_\pm(q^2), \quad T_\pm(q^2) = \tilde T_\pm(q^2).
\end{equation}
In~\cite{Jager:2012uw}, a parametrization of the following form was
suggested:
\begin{eqnarray}
  F(q^2) &=& F^{\infty}(q^2) +  a_F + b_F q^2/m_B^2 + {\cal
    O}([q^2/m_B^2]^2) . \label{eq:ffpara}
\end{eqnarray}
Here $F$ denotes any helicity form factor with $F^\infty(q^2)$
its HQ/LE limit, and the remainder the power corrections. More precisely,
$F^\infty(q^2)$ are functions (one for each form factor) among which
the HQ/LE relations hold, including perturbative corrections
\cite{Charles:1998dr,Beneke:2000wa,Beneke:2005gs}. The precise form of
$F^\infty(q^2)$ is ambiguous (see below) and defines a scheme for the
power correction terms. 

In \eq{eq:ffpara}, we have Taylor-expanded the power corrections
about $q^2=0$. As the form factors have 
no singularity in a circle of radius $m_{B_s}^2$ about the origin 
in the complex $q^2$ plane, this amounts to an expansion 
in the dimensionless ratio $q^2/m_{B_s}^2$ with coefficients of 
generic size $\mathcal{O}(\Lambda/m_B)$. In particular, 
the remainder term in \eq{eq:ffpara} 
should be a correction of a few percent throughout the low-$q^2$ 
region $q^2 < 6$ ${\rm GeV}^2$, and will be neglected in the following
(see also~\cite{Descotes-Genon:2014uoa}). We stress that beyond
this truncation there is no loss of generality in our decomposition.

The form factors obey a number of model-independent relations. First, there are
two exact constraints:
\begin{equation}
   T_+(0) = 0, \qquad S(0) = V_0(0).\label{eq:exactrel}
\end{equation}
Five further constraints hold in the HQ/LE limit, 
two to all orders\footnote{This was conjectured in~\cite{Burdman:2000ku},  
exhibited  at ${\cal O}(\alpha_s)$ in \cite{Jager:2012uw} and follows at 
${\cal O}(\alpha_s^2)$ from the results of \cite{Beneke:2005gs}. It seems
nevertheless clear that it is true to all orders~\cite{Beneke:2003pa}.}
in $\alpha_s$:
\begin{equation}
   V_+^\infty(q^2) = 0, \qquad T_+^\infty(q^2) = 0,     \label{eq:hqrel1}
\end{equation}
and three to zeroth order in perturbation theory only:
\begin{equation}
    V_-^{\infty}(q^2) = T_-^\infty(q^2), \qquad V_0^\infty(q^2) =
    T_0^\infty(q^2), \qquad  V_0^\infty(q^2) =  S^\infty(q^2) . \label{eq:hqrel2}
\end{equation}
The perturbative corrections to~\eq{eq:hqrel2} are unambiguously calculable
as convolutions of perturbative kernels with nonperturbative LCDA~\cite{Beneke:2000wa}.
Eqs.~\eq{eq:hqrel1} and~\eq{eq:hqrel2} reduce the nonperturbative
input in the exact HQ/LE limit from seven to two independent
functions, i.e., there are only two independent $F^\infty(q^2)$,
traditionally called the ``soft form factors.'' From
\eq{eq:hqrel2} it is clear that they correspond to the helicities -1 and 0, 
denoted $\xi_\perp(q^2)$ and $\xi_\parallel(q^2)$, respectively. 
Their $q^2$ dependence is not
calculable from first principles at present (the scaling in~\cite{Charles:1998dr}
is violated by radiative corrections \cite{Beneke:2000wa}); as a result,
not only the values at $q^2=0$ of the soft form factors but also their $q^2$-dependence
need to be modelled or determined experimentally.

At finite $m_B$, where the power corrections are non-zero, the soft form
factors are not uniquely determined, as the HQ/LE relations are invariant
under a shift:
\begin{equation}
   \xi_\perp(q^2) \to \xi_\perp(q^2) + f_\perp(q^2),
  \qquad
   \xi_\parallel(q^2) \to \xi_\parallel(q^2) + f_\parallel(q^2), \label{eq:ffredef}
\end{equation}
where $f_\perp$ and $f_\parallel$ are ${\cal O}(\Lambda/m_B)$.
It is customary to exploit this freedom by identifying $\xi_\perp(q^2)$
and $\xi_\parallel(q^2)$ with a pair of (finite-$m_B$) QCD form factors.

\begin{figure}[h]
\begin{center}
\includegraphics[scale=0.4]{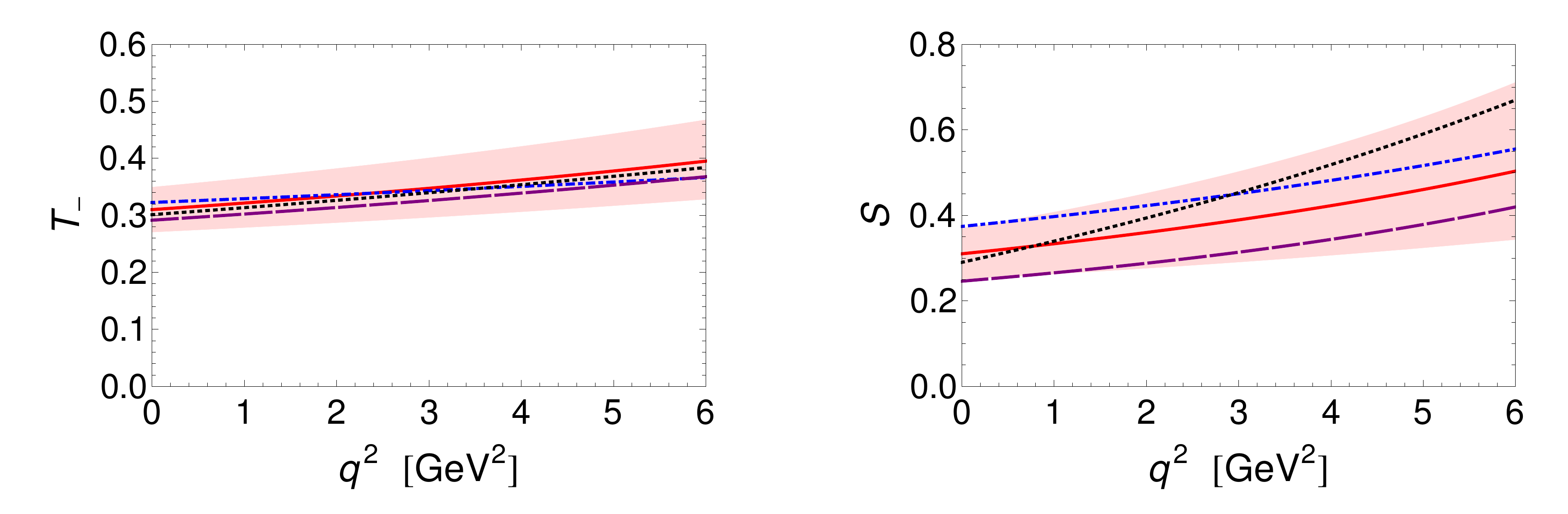}
\caption{The QCD form factors employed in this work to determine the soft form factors, $T_-(q^2)$ and 
$S(q^2)$ (red solid), compared to the determinations used as input in the LCSRs
~\cite{Ball:2004rg} (blue short-dashed), ~\cite{Khodjamirian:2010vf} (black dotted)
and Dyson-Schwinger equations~\cite{Ivanov:2007cw} (purple long-dashed) approaches.
Note that the error of $T_-(0)$ has been enlarged to cover the experimentally-driven determinations (see text).
\label{fig:QCDFFsinput}}
\end{center}
\end{figure}

In this work we use:
\begin{equation}
\xi_\perp(q^2)   \label{eq:xiperpdef}
    = T_1(q^2)
     = \frac{m_B}{2 |\vec k|}  T_-(q^2) - b_{T_+} \frac{q^2}{2m_B |\vec k|}
        + {\cal O}\left( \frac{\Lambda^2}{m_B^2}; \frac{\Lambda}{m_B}
                \left(\frac{q^2}{m_B^2} \right)^2 \right) ,
\hspace{0.7cm} \xi_\parallel(q^2)\equiv S(q^2),
\end{equation}
where $T_1(0)$ is a tensor form factor in the transversity basis.
 We parametrize the $q^2$-dependence by modifying the HQ/LE scalings of~\cite{Charles:1998dr}:
\begin{equation}
\xi_X(q^2)=\xi_X(0)\left(\frac{1}{1-q^2/m_B^2}\right)^{2+\alpha_X},\hspace{1cm}X=\perp,\parallel.
\label{eq:xiq2}
\end{equation}
The values at $q^2=0$:
\begin{equation}
\xi_\perp(0)=T_-(0)=0.31(4),\hspace{1cm}\xi_\parallel(0)=0.31(6),
\label{eq:xi0}
\end{equation}
are obtained as an average of results based on LCSR~\cite{Ball:2004rg,Khodjamirian:2010vf}
and Dyson-Schwinger equations~\cite{Ivanov:2007cw}; hereby the
errors are chosen such that the central values of these predictions
are covered, as is the value of  $T_-(0)$ suggested by the observed
$B \to K^* \gamma$ branching fraction, 
$T_-(0)\simeq
T_1(0)=0.277(13)$~\cite{Beneke:2001at,Altmannshofer:2008dz,Jager:2012uw}.~\footnote{The
  latter assumes that there is no NP in $C_7$ or $C_7^\prime$. The
  former Wilson coefficient is strongly constrained by inclusive $B \to
  X_s \gamma$ decay, and we assume it to be given by its SM value in
  this paper. The latter is constrained independently to be (in
  the present context) negligibly small by other
  observables as discussed and quantified in Sec.~\ref{sec:C7p}.}
We note that the identification $\xi_\perp(0)=T_-(0)$ implicitly fixes a
non-vanishing value of the $\mathcal{O}(\Lambda^2/m_B^2)$ residual
term in (\ref{eq:xiperpdef}). The parameters $\alpha_{X}$  in \eq{eq:xiq2} model 
the nonperturbative radiative violations to the ``naive'' HQ/LE 
scaling; we estimate them comparing 
again to the different calculations:
\begin{equation}
|\alpha_\perp|^{\rm max}=0.2,\hspace{1cm}|\alpha_\parallel|^{\rm max}=0.7.\label{eq:alphas}
\end{equation}
In Fig.~\ref{fig:QCDFFsinput} we show the $T_-(q^2)$ and $S(q^2)$ used in this
work compared to the central values of the calculations used as input. 
The error bands stem from the parameters $\xi_{X}(0)$ and $\alpha_{X}$,
that, together with the parameters in the $\alpha_s$ corrections~\cite{Beneke:2000wa}
to eqs.~\eq{eq:hqrel2} (see below), represent the only sources of
theoretical uncertainties that enter the helicity amplitudes through
the form factors in the HQ/LE limit.

While our errors on the soft form factors may seem to dwarf the
power corrections, this is a mirage: experimentally, one
studies observables for which the dependence on the soft
form factors cancels out if $\alpha_s$ and power corrections
are neglected. Hence the latter, parametrised by the coefficients
$a_F$ and $b_F$ in~\eq{eq:ffpara}, constitute a leading
source of uncertainty on these observables and must be carefully
considered. First, the exact 
relations~\eq{eq:exactrel} imply: 
\begin{equation}
a_{T_+} = 0,\qquad a_{V_0} = a_S. \label{eq:exactrelsas}
\end{equation}
On the other hand, the pairs of coefficients $a_{T_-}=a_{S}=0$ 
and $b_{T_-}=b_{S}=0$ since they are effectively absorbed 
in $\xi_{\perp,\parallel}(0)$ and $\alpha_{\perp,\,\parallel}$ 
respectively through the definitions
of the soft form factors detailed above. This, in turn,
also implies that $a_{V_0}=0$ using the second equation in \eq{eq:exactrelsas}.
Note that we cannot remove the remaining eight coefficients $a_{T_0}$, 
$b_{T_0}$, $a_{V_-}$, $b_{V_-}$, $a_{V_+}$, $b_{V_+}$, $b_{T_+}$ and $b_{V_0}$ 
while maintaining the heavy-quark relations \eq{eq:hqrel2}. 

These are dimensionless coefficients which represent a suppression
${\cal O}(\Lambda/m_B)$ over the values (for $a_F$)
and first derivatives (for $b_F$) of the form factors at $q^2=0$.
From the results in the HQ/LE limit we see that $F(0)\sim\xi_X(0)\simeq0.3$, 
while for the slopes eq.~\eq{eq:xiq2} leads to $dF(q^2)/dq^2|_{q^2=0}=\xi_X(0)(2+\alpha_X)/m_B^2$, 
which, taking into account eq.~\eq{eq:alphas}, can be numerically as large as $\sim1$ (in units of $m_B^{-2}$).
Thus, assuming for the moment that $\Lambda/m_B\sim0.10$, we find from power-counting 
arguments alone that:
\begin{equation}
|a_F^{\rm max,pc}|\simeq0.03,\hspace{1cm}|b_F^{\rm max,pc}|\simeq0.10. 
\label{eq:ab_pc}
\end{equation}
One should keep in mind that these estimates are \textit{ad-hoc} when 
interpreting the uncertainties derived from the bounds in \eq{eq:ab_pc}.
To be more precise, and except for their generic $\Lambda/m_B$ suppression 
we assume that the exact size (and sign) of the power corrections
are currently unknown.

\begin{figure}[h]
\begin{center}
\includegraphics[scale=0.4]{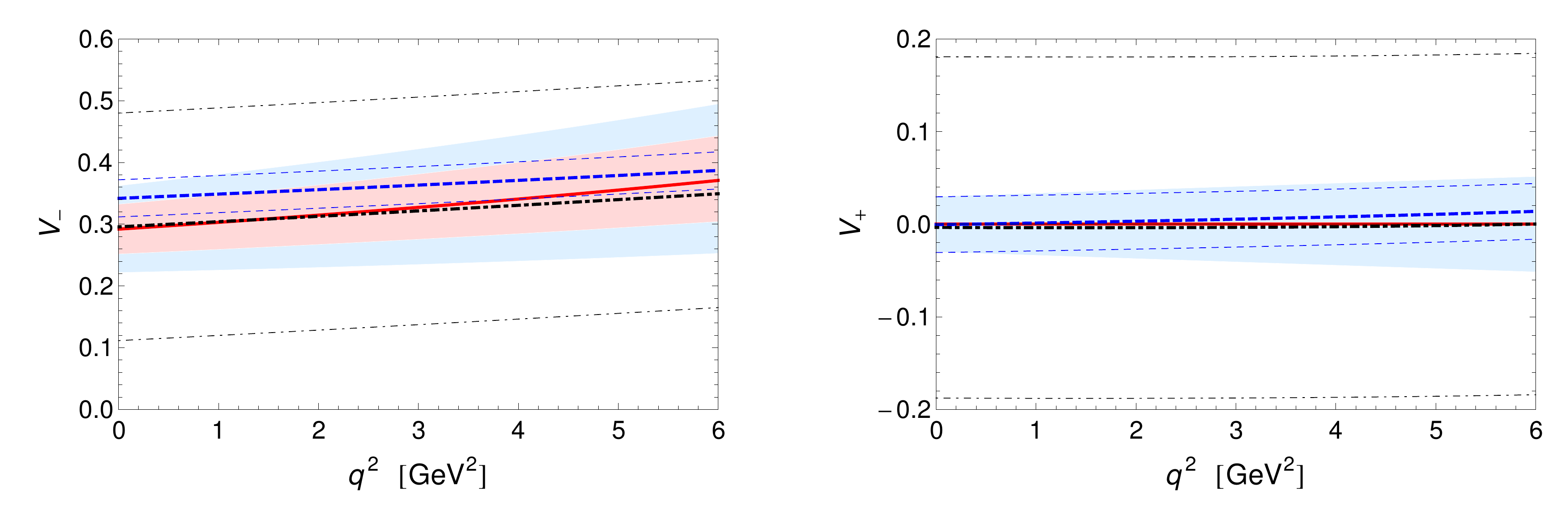}
\caption{The predictions on the form factors $V_-(q^2)$ and $V_+(q^2)$ obtained
in the QCDF approach described in this work. The blue band
are the uncertainties from power corrections added linearly to those from 
the errors of the parameters entering the perturbative corrections (red band). 
These are compared to results in two different LCSR calculations,
\cite{Ball:2004rg} (blue dashed) and \cite{Khodjamirian:2010vf} (black dot-dashed),
where the thinner lines are the errors obtained ignoring correlations
and taking the maximum possible value when transforming to the helicity
basis used here.
\label{fig:QCDFFsvpvm}}
\end{center}
\end{figure}

One can also use the model-independent parametrization in 
eq.~\eq{eq:ffpara} to implement further constraints that could be obtained from 
first-principles in QCD or to build in nonperturbative calculations of the 
form factors and test them (see e.g. \cite{Descotes-Genon:2014uoa}). 
This discussion is of the utmost importance in instances 
where experimental tensions with the SM appear in
observables that are not only sensitive to short-distance Wilson coefficients
but also to the power corrections to the form factors. 

In connection to this, and as an illustration that will become useful 
for the phenomenological discussion below, we show in Fig.~\ref{fig:QCDFFsvpvm}
the vector form factors $V_\pm(q^2)$ used in this work together with two 
different predictions from LCSRs. It is interesting to note at this point that the 
calculation in~\cite{Ball:2004rg} implies a power correction to $V_-(q^2)$ which is
consistent with the power-counting estimate in~\eq{eq:ab_pc} but favouring
an overall \textit{positive} sign, i.e.
\begin{equation}
\frac{V_-(0)}{T_-(0)}\gtrsim\frac{V_-^\infty(0)}{T_-^\infty(0)}. 
\end{equation}

Finally, let us also remark that the identifications 
used in eq.~\eq{eq:xi0} are arbitrary and other QCD form factors could have been 
employed in their stead. This amounts to re-arrangements of power corrections, cf. eq~(\ref{eq:ffredef}), 
so that comparing different schemes allows to test the robustness of the approach
introduced in~\cite{Jager:2012uw} and refined here.
Our particular choice is 
different to the one taken in~\cite{DescotesGenon:2011yn,DescotesGenon:2012zf}
which is based on using the vector form factors, i.e. $V_-(q^2)$ and $V_0(q^2)$
in the helicity basis. We prefer to retain the tensor form factor $T_-(q^2)$ because
its value at $q^2=0$  can be determined from experimental data~\cite{Beneke:2000wa,Jager:2012uw}.
The form factor $S(q^2)$ appears in the amplitude only multiplied
by lepton masses, and it
is argued in ref.~\cite{Descotes-Genon:2014uoa} that using $V_0(q^2)$ to fix
$\xi_\parallel(q^2)$ in its stead could reduce the impact of the power corrections
in the predictions. However, eq.~\eq{eq:exactrelsas} implies that the power
corrections to $V_0(q^2)$ enter only through $b_{V_0}$ which has a marginal
effect in the total uncertainty.  We will explicitly demonstrate
 this by comparing both choices in the phenomenological discussion below.

\subsubsection{Nonfactorizable term}
\label{sec:theononfact}

The matrix element of the hadronic weak Hamiltonian, affecting 
only the three vector helicity amplitudes through the $h_\lambda$ terms in \eq{eq:HV},
can also be split into a HQ/LE limit and a power correction term,
\begin{equation}
  h_\lambda(q^2) = h_\lambda^\infty(q^2) + r_\lambda(q^2). \label{eq:nonlocal0}
\end{equation}

The leading-power term $h_\lambda^\infty$ can be calculated systematically to
any order in $\alpha_s$ in QCD factorization~\cite{Beneke:2001at}. 
It carries a well-defined $q^2$-dependence. 
In particular, to ${\cal O}(\alpha_s^0)$ it amounts to the well
known substitutions $C_7 \to C_7^{\rm eff}$ and $C_9 \to C_9^{\rm
  eff}(q^2)$ in \eq{eq:HV} and the addition of a single, CKM-suppressed annihilation diagram.

\begin{figure}[h]
\begin{center}
\includegraphics[scale=0.4]{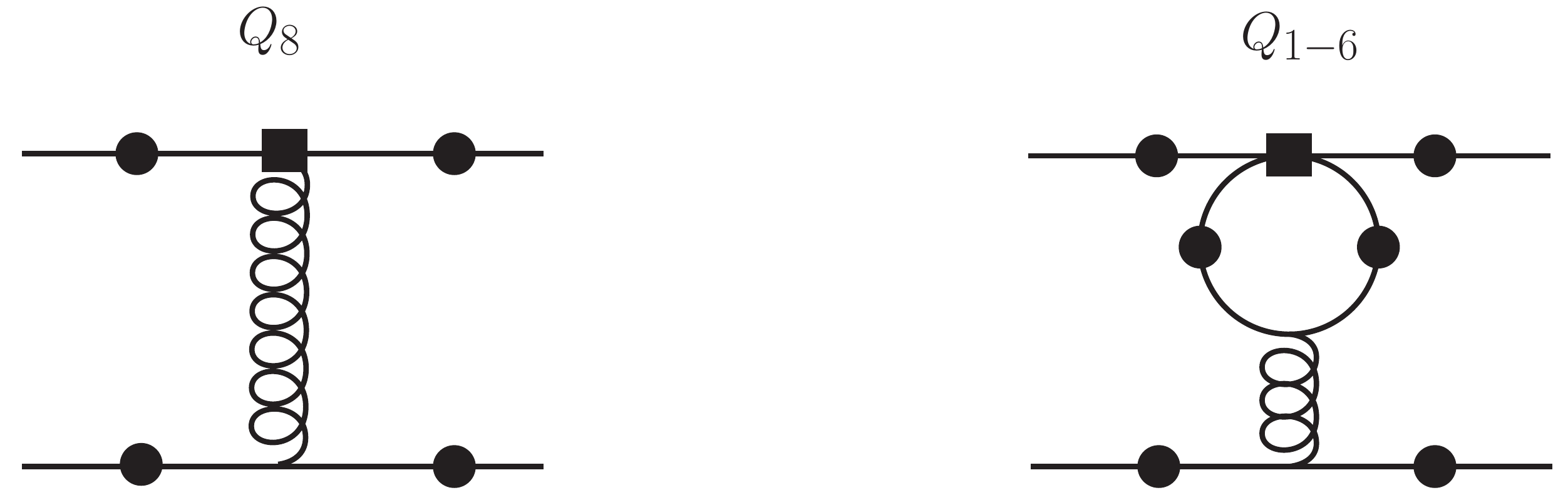}
\caption{Spectator scattering diagrams for $B \to V
  \gamma^{(*)}$. The bullets denote the possible photon attachments.
\label{fig:specscatt}}
\end{center}
\end{figure}

The power-correction terms $r_\lambda$ are more complicated than
in the form factor case. The hadronic weak hamiltonian comprises two
operators $Q_1^c$, $Q_2^c$ involving a charm quark pair, as well as
four-quark operators containing light quarks and the chromomagnetic
penguin operator. Of these, the charmed operators come with large CKM
and Wilson coefficients, presumably
giving the most important contributions that we will denote by $r^c_\lambda$.

A key conclusion in \cite{Jager:2012uw} was that while
  $r^c_\lambda$ is \textit{not} negligible for $\lambda=-,0$, it
respects the same helicity hierarchy as the factorizable terms, i.e.\
$|r^c_+| \ll |r^c_-| , |r^c_0|$. This relied on a LCSR estimate of
soft gluon emission from the charm loop, and we clarify here the
precise relation to the QCD factorization result in the
heavy-quark limit.

Recall first that the HQ/LE limit is given in terms of Feynman diagrams
such as those shown in Figure \ref{fig:specscatt}, computed for soft
``constituents'' of the $B$-meson and collinear ones of the $K^*$,
convoluted with leading-twist light-cone distribution amplitudes.
 Schematically,
\begin{equation}
   h_\lambda = \int_{0}^1 du \,\phi_{K^*}(u) T(u, \alpha_s) + {\cal O}(\Lambda/m_b) ,
\end{equation}
where $u$ is the fraction of the $K^*$ momentum carried by one
of the $K^*$ constituents and dependence on the $B$-meson constituent
momenta has been suppressed.
The internal lines in the graphs can have hard [${\cal O}(m_b^2)$] or
hard-collinear [${\cal O}(m_b \Lambda)$] virtualities. A Wilsonian
picture is provided (just as for the form factor case)
by SCET, whereby in two
matching steps the hard and the hard-collinear degrees of freedom are
integrated out, leaving a theory with only ${\cal O}(\Lambda^2)$
virtualities, where soft and collinear modes no longer talk to each
other and $T(u, \alpha_s)$ is a Wilson coefficient.
In fact the picture is not quite complete, as some convolutions are
not convergent: these, however, can be absorbed into heavy-to-light
form factors, times so-called vertex corrections.
There are also  annihilation graphs that converge for
  $\lambda=\pm1$ for all $q^2$.
A necessary and sufficient
requirement for convergence is a sufficiently fast fall-off of the
LCDA at the endpoint, which holds at leading twist. 
This ensures that the ``end-point'' contributions to
the convolutions, where the collinear ``constituent'' really is soft,
and the LCDA formalism does not apply, is
suppressed by at least one power of $\Lambda/m_B$ \cite{Beneke:2000ry}.

Further sources of power corrections arise from similar graphs
  (with two $K^*$ ``constituents'') convoluted with higher-twist
  two-particle LCDAs, and from graphs with more constituent lines
  convoluted with multi-particle LCDAs (also of higher twist).
Sometimes, such higher-twist contributions exhibit endpoint divergences.
Note that at the (spectator) endpoint the virtualities of the
internal lines are reduced. For example, the gluon in Fig.\
\ref{fig:specscatt} is generically hard-collinear, but soft in the
endpoint region, if the photon does not attach to the spectator
line. One can introduce a cutoff $\Lambda_h$
separating collinear from soft momenta. Then schematically,
\begin{equation}
   r^c_\lambda \sim \int_{\Lambda_h}^1 du \,\phi_{K^*}(u) T(u, \alpha_s) +  r^c_{\lambda,\rm soft} .
 \end{equation}
The first term represents the contributions involving hard-collinear
gluon exchanges and is calculable in perturbation theory (albeit
cutoff dependent). The second contribution cannot be computed in QCD
factorization, however it can be represented by an operator matrix
element $\langle K^*  | O_{\bar s G b} | B \rangle$ where
$O_{\bar s G b}$ is an operator involving one soft gluon
field. The endpoint contribution in this case would not have a
  relative power suppression; this can again be argued based on the
  endpoint behaviour of the LCDA and kernel \cite{Beneke:2000ry}, or
  simply on the grounds that, even though not calculable,
it needs to cancel the cut-off dependence.

 At the
two-particle LCDA level,  the hard-collinear term for an insertion of the $(V-A) \times (V-A)$
operators $Q_1^c, Q_2^c$ cannot generate an $r^c_+$ amplitude because
of chirality conservation in QCD (if light-quark masses are neglected)
\cite{Burdman:2000ku,Beneke:2000wa}. It is clear that this also
  applies to the twist-3 (first subleading power) two-particle
  contribution.
There are also twist-three three-particle contributions,
corresponding to an extra outgoing collinear gluon line. They do
not vanish by the $V-A$ structure of weak interaction, as the extra
gluon field can carry helicity $+1$. These three-particle twist-three
contributions have not been computed in the
literature.\footnote{A subset of twist-3 two-particle contributions
  relevant to the isospin asymmetries in $B \to K^* \gamma$ and $B \to
  K^* l^+ l^-$ has been calculated in
  \cite{Kagan:2001zk,Feldmann:2002iw}.}
In LCSR computations of the $B \to V$ form factors,
they give small contributions \cite{Ball:1998kk}. For the
present case, we  note that the hard-collinear contributions are suppressed by a factor
of $\alpha_s$, and it appears clear that they are endpoint convergent
from the fact that the three-particle twist-three vector meson LCDA
vanishes linearly at the quark and antiquark endpoints (and
quadratically at the gluon endpoint) \cite{Ball:1998sk}.

In summary, all hard-collinear contributions are at least
$\Lambda^2/m_B^2$ or $\Lambda/m_B \times \alpha_s$ suppressed,
and the same is true for any contribution involving a collinear gluon,
including if extra soft gluons are radiated. Hence the fate 
of $r^c_+$ is determined (at order $\Lambda/m_B$) by purely soft gluon
emission from the charm loop, while $r^c_-$, $r^c_0$ receive
small corrections to their nonvanishing leading-power values from this mechanism.
For  $q^2 \ll 4 m_c^2$, 
 the interaction of the charm loop with a soft gluon background
  can be represented through a series of light-cone
operators with matrix elements scaling as
$[\Lambda^2/(4 m_c^2)]^n$, $n=1, 2, \dots$
\cite{Khodjamirian:2010vf}.\footnote{We stress that the gluon is
    soft in the $B$ rest frame. This is different from the case where one
considers extra collinear gluons, such as in deriving a
light-cone sum rule with $K^*$ distribution amplitudes; see the
discussion in \cite{Jager:2012uw}. However,
as explained, it is soft gluon emission that determines power
corrections incalculable in QCD factorization.}
The  $B \to K^*$ matrix element of the leading (1-soft-gluon) term in this
operator expansion provides an approximation of $r^c_{\lambda, \rm
  soft}$, accurate up to $\Lambda^4/(4 m_c)^4)$ corrections.
We have verified that the result in \cite{Khodjamirian:2010vf}
corresponds to the identification $\Lambda m_B \sim 4 m_c^2$.
One obtains generic sizes of $\Lambda/m_B$ for $r^c_{\lambda, \rm soft}$, as expected. For
$q^2=0$ this was indeed the conclusion reached already in
\cite{Grinstein:2004uu}.

To go further in estimating the hadronic matrix elements,
in \cite{Jager:2012uw} we adapted a LCSR developed in
\cite{Khodjamirian:2010vf} to the helicity amplitudes
$r^c_\lambda$. LCSRs equate the matrix element of a
perturbative two-point correlator to a sum over a complete set of
hadronic states, which is then truncated to its first term by means of
a duality threshold model. This first term contains the desired
hadronic matrix element. In the case at hand, the correlation
function vanishes for $\lambda=+1$ up to terms suppressed by an extra factor of
$\Lambda/m_B$ \cite{Jager:2012uw}. Barring systematic cancellations between the different
terms in the hadronic sum, for which we cannot identify a mechanism,
this implies that $r^c_+ = {\cal O}(\Lambda^2/m_B^2)$. No such
suppression takes place for $\lambda=-, 0$, which are ${\cal O}(\Lambda/m_B)$.

Based on these considerations, we parametrize the remainder term as:
\begin{eqnarray}
   r^c_\lambda(q^2) &=& A_\lambda
      + B_\lambda \frac{q^2}{4 m_c^2} \label{eq:hpara} ,
\end{eqnarray}
where $A_\lambda$ and $B_\lambda$ are constants of order $\Lambda^2/(4m_c^2)$.
This generalizes the parametrization in \cite{Jager:2012uw}, where
$r_\lambda(q^2)$ was approximated by its $q^2=0$ value, equivalent to
a shift in the Wilson coefficient $C_7$ in \eq{eq:HV}
and appropriate for the phenomenological
discussion for $q^2 \le 3$ GeV${}^2$, a restriction we do not wish to
make here. The $B_\lambda$-terms parametrize the leading corrections
to a pure $1/q^2$-dependence once the photon propagator is taken into
account, and give rise to a $q^2$-independent contribution to the
amplitude, and an estimate of the uncertainty due to long-distance
charm contributions away from the kinematic endpoint.
Attempts to improve on this by including higher powers of
$q^2/(4 m_c^2)$ appear futile, as those will only become relevant once
$q^2/(4 m_c^2) \sim 1$ and the expansion breaks down. 

Concerning the numerical values of $A_\lambda$ and $B_\lambda$,
  we allow arbitrary complex phases for them. This is because, even when
  restricting to the operators $Q_1^c$ and $Q_2^c$, $r_\lambda$ is
  affected by ``charmless'' multiparticle cuts, just as happens already
  in the leading-power, perturbatively calculable charm loop result
  \cite{Beneke:2001at}. (See
  also the discussion in terms of hadronic final state interactions in 
  \cite{Khodjamirian:2012rm}.)
 Although this means that the $q^2$ dependence is not analytic in any disc around
  the origin, the dependence on the charm mass should still be regular
  and admit an expansion in the variable $q^2/(4 m_c^2)$ (with
  $m_c$-independent, non-analytic coefficients), as the charm quark
  pair is always off shell. In practice
  allowing an arbitrary strong phase has little
  effect on the uncertainties on the observables considered below.

\begin{figure}[h]
\begin{center}
\includegraphics[scale=0.45]{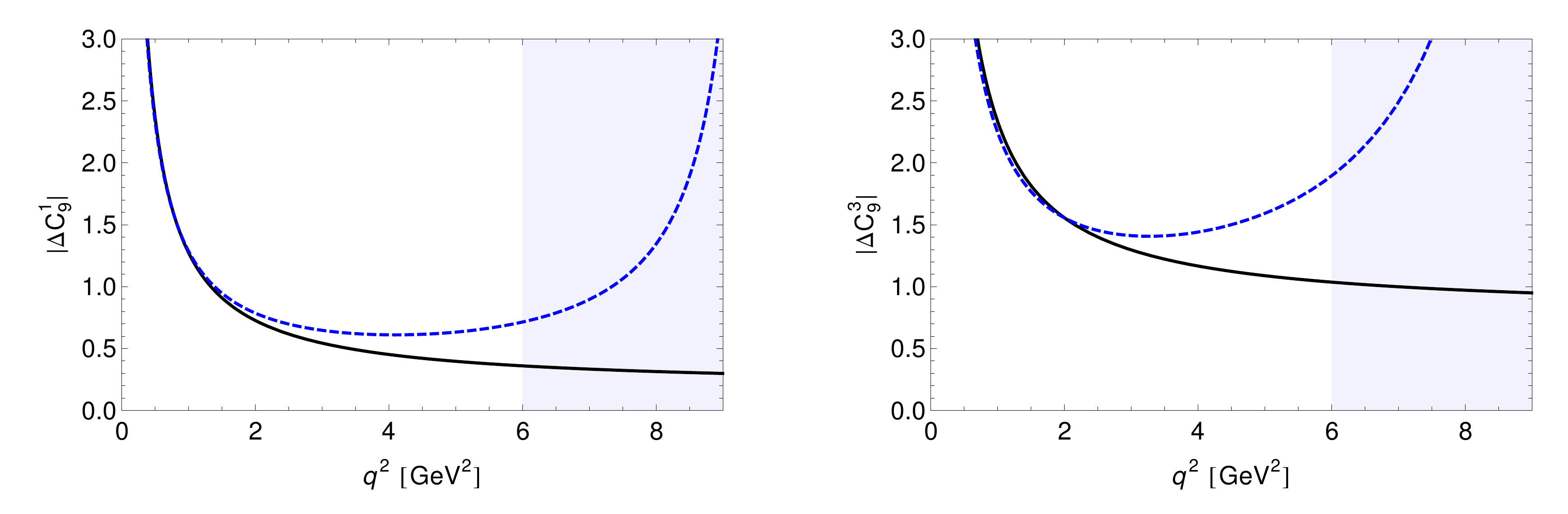}
\caption{
Long-distance charm contributions parametrized as a correction 
to the Wilson coefficient $C_9$. The (black) solid line results from 
the maximum possible contribution as shown in ref.~\cite{Khodjamirian:2010vf}.
We show the modulus of the (complex) correction extrapolated up 
to $q^2\simeq 6$ GeV$^2$. The (blue) dashed lines correspond to the
hadronic parametrization used in~\cite{Khodjamirian:2010vf}.
\label{fig:C9cc}}
\end{center}
\end{figure}
It is instructive to represent $r^c_\lambda$ as a
  (helicity-dependent) shift to $C_9^{\rm eff}(q^2)$.
In Fig.~\ref{fig:C9cc} we show the long-distance charm contributions
used in this work parametrized as a correction to the Wilson coefficient $C_9$.
The (black) solid line is the result of a fit of
eq.~\eq{eq:hpara} to the upper error band shown in Fig. 5 of
ref.~\cite{Khodjamirian:2010vf} that is the result
of the LCSR calculation valid up to $q^2\simeq 4$ GeV$^2$. We show the
modulus of the (complex) correction and extrapolate up to 
$q^2\simeq 6$ GeV$^2$. More precisely,~\eq{eq:hpara} corresponds to a
parametrization of the type  $|\Delta C_9^i|=2\,m_b\,m_B/q^2\,\delta_{i1}+\delta_{i2}$,
and the fits give $\delta_{11}=0.02$, $\delta_{12}=0.18$, $\delta_{31}=0.03$, 
$\delta_{32}=0.78$. This correction is implemented numerically in 
this work as a flat error in the corresponding helicity amplitudes.
The contribution $r_+$ is additionally power suppressed as discussed
above, and we will use the same parameters as in $r_-$
but multiplied by $\Lambda/m_B$.   

In Fig.~\ref{fig:C9cc} we also show the parametrization  
used in~\cite{Khodjamirian:2010vf} as (blue) dashed lines, where the
partonic result is matched to a hadronic representation via a 
dispersion integral and a model for the continuum contribution.
This is shown as an illustration of the possible large long-distance
effects that could be induced by the $Q_1^c$ and $Q_2^c$ operators at
$q^2\gtrsim6$ GeV$^2$ and whose description requires introducing
model-dependence. This is also consistent with the findings of 
ref.~\cite{Lyon:2014hpa}.

Other contributions to $r_\lambda$ can also be investigated. Those 
induced by the chromomagnetic penguin operator $Q_8$ have been 
studied in the context of LCSR in~\cite{Khodjamirian:2012rm} and
~\cite{Dimou:2012un}, an their contributions turn out to be very small.
The contributions involving light-quark loops can be problematic at low $q^2$
since their treatment in QCDF is the dual to the one induced by light
vector resonances. However, they always come doubly CKM suppressed or multiplied
by small Wilson coefficients. A study of the impact of the duality violation
(in relation to the QCDF result) was done using vector-meson dominance 
in~\cite{Jager:2012uw} and it turned out to be negligibly small in
the binned angular observables. It was also shown 
that $r_+^{u,d,s}$ for the light quarks is also suppressed by $(\Lambda/m_B)^2$.

For all this, we neglect the power corrections to the other terms, effectively
absorbing them into $r_\lambda^c$ and will treat all the corrections to $r_+$
suppressed by $(\Lambda/m_B)^2$.

\section{Angular observables and the analysis of the experimental data}
\label{sec:phenogen}
The $q^2$-dependent angular distribution (summed over lepton spins)
is quadratic in the helicity amplitudes and has been given in~\cite{Jager:2012uw}.
Certain ratios of angular coefficients are favoured because of their
reduced sensitivity to form factors. In particular, we will discuss the
so-called $P_i^{(\prime)}$ basis which was introduced 
in~\cite{Matias:2012xw,DescotesGenon:2012zf}. This is an exhaustive set
of observables, constructed from ratios of the angular coefficients and
engineered to cancel most of the hadronic uncertainties in the HQ/LE 
limit.

In order to illustrate this and critically re-examine the
residual uncertainties on those observables, we will focus on
two of them, called $P_1$ and $P_5'$ in \cite{Matias:2012xw,DescotesGenon:2012zf}.
In terms of the helicity amplitudes, they read:
\begin{eqnarray}
   P_1  &=& 
        \frac{  -2\,{\rm Re} (H_V^+ H_V^{-*} + H_A^+ H_A^{-*} ) }
                {  | H_V^+|^2 + |H_V^-|^2 + |H_A^+|^2 + |H_A^-|^2  } , \\
   P_5'  &=&
        \frac{ {\rm Re} [(H_V^- - H_V^+) H_A^{0*} +
                                                       (H_A^- - H_A^+) H_V^{0*} ] }
                {\sqrt{(|H_V^0|^2 + |H_A^0|^2) 
                        (| H_V^+|^2 +|H_V^-|^2 +|H_A^+|^2 +|H_A^-|^2) }}
\end{eqnarray}
where we have neglected the muon mass for clarity and have introduced
the short-hand notation $H_{V,A}(\lambda)=H^\lambda_{V,A}$.

In certain approximations $P_1$ and $P_5'$ become free of
nonperturbative uncertainties. In the HQ/LE limit and
neglecting $\alpha_s$ corrections, as well as the contributions
$h_\lambda$ from the hadronic weak Hamiltonian, the 
$\lambda=+$ helicity amplitudes vanish and $V_\lambda(q^2) = T_\lambda(q^2)$.
As a result, in these limits and in the SM~\footnote{We will ignore
in this discussion the strange quark mass which produces
an effect suppressed by $m_s/m_b$ in $P_1$.},
\begin{eqnarray}
 P_1  &=& 0,   \label{eq:p1clean}  \\
 P_5'  &=&  \frac{ {\rm Re}[C_{10}^*C_{9,\perp}+C_{9,\parallel}^*C_{10}]}
    {\sqrt{ ( |C_{9,\parallel}|^2 + |C_{10}|^2)
           (|C_{9,\perp}|^2 + |C_{10}|^2) } } , \label{eq:p5pclean}
 \end{eqnarray}
where 
$ C_{9,\perp}= C_9^{\rm eff}(q^2) + \frac{2\, m_b\, m_B}{q^2}
C_7^{\rm eff}$, 
$ C_{9,\parallel} = C_9^{\rm eff}(q^2) + \frac{2\,m_b}{m_B} C_7^{\rm eff}$,
and the $P_i^{(\prime)}$ are functions of the Wilson coefficients 
alone. 

Thus, the leading sources of uncertainties for the observables in
the $P_i^{(\prime)}$ basis are due to the presence of nonfactorizable 
contributions as well as to corrections to the HQ/LE form factor
relations. To see this explicitly, note that
\begin{eqnarray}
  P_5' &=& P_5'|_{\infty} \Bigg(1
      + \frac{a_{V_-} - a_{T_-} }{\xi_\perp} \frac{m_B}{|\vec k|}\frac{m_B^2}{q^2} C_7^{\rm eff}
              \frac{C_{9,\perp} C_{9, \parallel} - C_{10}^2 }{(C_{9,\perp}^2 + C_{10}^2) (C_{9,\perp} +
                      C_{9,\parallel} ) }
\nonumber \\
 &&    \qquad \qquad + \frac{a_{V_0} - a_{T_0} }{\xi_\parallel}\, 2\,C_7^{\rm eff}
              \frac{C_{9,\perp} C_{9, \parallel} - C_{10}^2 }{(C_{9,\parallel}^2 + C_{10}^2) (C_{9,\perp} +
                      C_{9,\parallel} )  }
\nonumber \\
 &&      \qquad \qquad + \,8\pi^2\, \frac{\tilde h_-}{\xi_\perp}\frac{m_B}{|\vec k|} \frac{m_B^2}{q^2}
       \frac{C_{9,\perp} C_{9,\parallel} - C_{10}^2}{C_{9,\perp} +
                      C_{9,\parallel}  }
   + \mbox{further terms} \Bigg) + {\cal O}(\Lambda^2/m_B^2),
\end{eqnarray}
where for simplicity we have assumed real Wilson coefficients,
$\tilde h_+$ denotes the nonlocal term $h_\lambda$ with its
leading term removed (absorbed into $C_9^{\rm eff}$), and we have
neglected the difference between $m_b$ and $m_B$ as a higher-order effect.
We see in the second  term on the first line the presence of the power correction
combination $a_{V_-} - a_{T_-}$. This is invariant under change of
soft form factor scheme [cf. \eq{eq:ffredef}] -- in particular it does not matter whether $V_-$ or $T_-$
is identified with $\xi_\perp$, implying $a_{V_-} = 0$ or $a_{T_-}
=0$, respectively.  Similarly, power corrections to the helicity-zero
form factors enter only in the combination ($a_{V_0}$ - $a_{T_0}$)
(second line).
Both can be understood by observing that form factors cancel out of
$P_5'$ completely if  $C_7^{\rm eff}$, the $\lambda=+$ amplitudes, and nonfactorizable
corrections are all neglected. As a result, form factor uncertainties enter only through
interference of the tensor and vector form factors, and of form
factors and nonfactorizable corrections.
Put another way, form factor uncertainties enter the $P_i$
only through deviations of the form factor differences
$T_\lambda - V_\lambda$ from their zero HQ/LE limit.
(For $\lambda = -,0$, these could be traded for the deviations of
the ratios $T_\lambda/V_\lambda$ from one.)
At the model-independent
level, lacking first-principles results on the individual form factors
all one can achieve is to parameterise the power correction
$T_\lambda - V_\lambda$ in some way. For example, without loss of
generality one can parameterise
$$
 T_-(q^2) - V_-(q^2)   \equiv  a_{T_-} + b_{T_-} \frac{q^2}{m_B^2} + \cdots
       \qquad \mbox{(corresponding to } a_{V_-} = 0 )
$$
or one can take
$$
  T_-(q^2) - V_-(q^2)  \equiv  - a_{V_-} - b_{V_-} \frac{q^2}{m_B^2} + \cdots
       \qquad \mbox{(corresponding to } a_{T_-} = 0 ) .
$$
This is an equivalent way of defining the same power correction
schemes discussed above.
Clearly, any model prediction such as a sum rule calculation can be
expressed in either parameterisation. Of course, the prediction for a physical
observables does not depend on which choice is made, as the scheme
dependence of the model calculation cancels against that in the
observable calculation (in the present case, simply $a_{T_-}$ and
$-a_{V_-}$ are traded for one another in the intermediate stages, and
similarly
$b_{T_-}$ and $-b_{V_-}$).

The interference is most
important if $C_7^{\rm eff}$ and $C_9^{\rm eff}(q^2)$ are comparable,
as happens in particular around the zero-crossing of $P_5'$.
The term displayed on the last line involves nonfactorizable
corrections. All three terms demonstrate how the soft form
factors with their associated uncertainties re-enter at subleading power.
The full expression is quite lengthy and depends on all
power-correction parameters and the three nonlocal terms.
A similar sensitivity to power corrections occurs in most of the other
angular coefficients, and in the observables in the $P_i^{(\prime)}$ basis
built from them. This includes the locations of the
zero-crossings of these observables.

In striking contrast, the ${\cal O}(\Lambda/m_B)$ power corrections to
$P_1$ take the simple form
\begin{eqnarray}
P_1 &=& \frac{1}{C_{9,\perp}^2 + C_{10}^2}\, \frac{m_B}{|\vec k|}\Bigg(
                   - \frac{a_{T_+}}{\xi_\perp} \frac{2\,m_B^2}{q^2} C_7^{\rm eff} C_{9,\perp} 
                   - \frac{a_{V_+}}{\xi_\perp} \,(C_{9,\perp} C_9^{\rm eff} + C_{10}^2)
                   - \frac{b_{T_+}}{\xi_\perp} \,2 C_7^{\rm eff} C_{9,\perp}
\nonumber \\
&&              - \frac{b_{V_+}}{\xi_\perp}\,\frac{q^2}{m_B^2} (C_{9,\perp} C_9^{\rm eff} + C_{10}^2)
                   + 16\pi^2  \frac{h_+}{\xi_\perp}\, \frac{m_B^2}{q^2} C_{9, \perp} 
\Bigg) +  {\cal O}(\Lambda^2/m_B^2) .
\end{eqnarray}
Apart from  depending on only one soft form factor and fewer power-correction and
non-local parameters, these terms suffer further suppression:
$a_{T_+}$ vanishes exactly as discussed in\ref{sec:theoff}, the next three terms are suppressed by a
power of $q^2/m_B^2$ relative to the denominator at small $q^2$, and
$h_+$ has an extra power suppression as discussed in the previous section.
As a result, $P_1$ vanishes like ${\cal O}(\Lambda^2/m_B^2), {\cal
  O}(C_9^{\rm eff}/C_7^{\rm eff} \times q^2/m_B^2 \times \Lambda/m_B)$
at small $q^2$ in the SM. By contrast, in the presence of non-zero
$C_7'$ it is order one. Analogous is the case of the
CP-asymmetry $P_3^{CP}$~\cite{Becirevic:2011bp,Jager:2012uw},
which at low $q^2$ cleanly probes a BSM weak phase in $C_7^{\prime}$.

Finally let us comment on the extension of our conclusions to the observables
in the $S_i$ basis, where the total decay rate is used to normalize the 
angular observables~\cite{Altmannshofer:2008dz}. The theoretical
errors are in general expected to be larger than in the
$P_i^{(\prime)}$ basis, because the sensitivity to $\xi_\perp(q^2)$ and
  $\xi_\parallel(q^2)$ is  enhanced in the heavy-quark limit without
  the $P_i'$ basis optimisation. However, the suppression in the SM of the
$\lambda=+$ amplitudes still implies the same ${\cal O}(\Lambda^2/m_B^2)$,
${\cal O}(C_9^{\rm eff}/C_7^{\rm eff} \times q^2/m_B^2 \times \Lambda/m_B)$
suppression seen above for the ``un-optimized'' versions of
$P_1$ and $P_3^{CP}$, $S_3$ and $A_9$ respectively, and the additional
form factor uncertainties affect only the normalisation of the
residual term in these two special cases, which is very small at small
$q^2$.

\subsection{Statistical framework and predictions in the SM}
\label{sec:stats}

In the analysis of experimental data one must specify
the treatment of the theoretical uncertainties in
the statistical framework to be used. A frequentist scheme that has been
successfully applied to the analysis of the CKM unitarity
triangle by the \textit{CKMfitter} collaboration is the
range fit (or $R$fit) method~\cite{Hocker:2001xe}.
In this approach, the $\chi^2$ is first constructed in the usual way, 
based on a vector of experimentally measured observables $\vec{x}$
with experimental uncertainty $\vec{\sigma}$.~\footnote{The correlations for the
data used in this paper have not been published and we assume in our fits
that the different measurements are uncorrelated.} The theoretical
determination of the observables depends on two types of variables:
\textit{(i)} A set $\vec{C}$ of short-distance Wilson
coefficients of the effective weak Hamiltonian; \textit{(ii)}
hadronic parameters, $\vec{y}$ that can be determined using
various nonperturbative methods with some systematic uncertainty 
$\vec{\delta}$. A piece is then added to the $\chi^2$ 
that does not contribute unless any of the components $y_i$
leaves the range determined by its uncertainty, $\delta_i$.
More explicitly:
\begin{eqnarray}
 \chi^2(\vec{C},\vec{y})=\left\{\begin{array}{c}
 \sum_i\frac{\left(x_i-x^{\rm th}_i(\vec{C},\vec{y}\,)\right)^2}
 {\sigma_i^2},\hspace{0.5cm} {\rm if}\hspace{0.4cm} y_{k}\in[\bar{y}_k-\delta_k,\,\bar{y}_k+\delta_k]\hspace{0.4cm}\forall k \\
  \infty, \hspace{0.5cm} {\rm otherwise}\end{array}\right..
\end{eqnarray}
Once it is in this form, one can treat the set $\vec{y}$ 
as nuisance parameters to construct a ``profile $\chi^2$'' 
depending on the Wilson coefficients alone:
\begin{eqnarray}
\tilde{\chi}^2(\vec{C})=\underset{\vec{y}}{\rm min}\,\chi^2(\vec{C},\vec{y}), 
\end{eqnarray}
and determine confidence level (CL) intervals for the $C_i$ in a frequentist
fashion and irrespective of the values of the other QCD parameters.
Obviously, QCD parameters with particular interest, e.g. for testing 
nonperturbative calculations, can be promoted to the set of ``interesting''
parameters in $\vec{C}$ and be determined simultaneously from the experimental
data. 

Another statistical framework would be to implement the theoretical
uncertainties in a $\chi^2$ by adding in quadratures to
the experimental uncertainties~\cite{Descotes-Genon:2013wba,Altmannshofer:2014rta,Hurth:2013ssa}.
This implicitly assumes that the theoretical errors distribute normally or
can be measured independently, although this will be rarely the case. 

These considerations have to be kept in mind when interpreting the 
theoretical predictions of the observables. In particular, it is clear that 
the $R$fit method scans the space of parameters $\vec{y}$ within the interval defined 
by its uncertainty, searching for CL intervals that maximize the agreement 
between theory and experimental data. Thus, our uncertainties must be interpreted
in terms of the maximal spread of theoretical predictions that will be considered in the global
fits. On the other hand, it is important to stress that the nuisance parameters 
are also implicitly fitted and the range of their values will be further constrained
in the analysis.

\begin{table}[h]
\centering
\caption{Results for the bin $[1,\,6]$ GeV$^2$ in the SM for a selection
of observables and using different schemes for the estimation of the
theoretical uncertainties. We compare with independent calculations in the
literature~\cite{Beaujean:2012uj,Beaujean:2013soa,Descotes-Genon:2014uoa} whenever it is 
possible. In the last column we show the experimental data.\label{tab:SMpredsOld}}
\begin{tabular}{|c|c|c|c||c|c||c|}
\hline
&Max. Spread &  Max. spread ($V_-$ and $V_0$)&1$\sigma$ Gaussian& Ref.~\cite{Beaujean:2012uj,Beaujean:2013soa} & Ref.~\cite{Descotes-Genon:2014uoa}& Expt.\\
\hline
$P_1$&$-0.03^{+0.22}_{-0.24}$&$-0.03^{+0.22}_{-0.23}$&$-0.03^{+0.13}_{-0.13}$&$--$&$0.009^{+0.038}_{-0.044}$&$0.15(0.40)$\\
\hline
$P_2$&$-0.12^{+0.41}_{-0.37}$&$-0.11^{+0.37}_{-0.33}$&$-0.10^{+0.19}_{-0.19}$&$-0.15^{+0.07}_{-0.07}$&$-0.21^{+0.17}_{-0.20}$&$-0.66(23)$\\
\hline
$P_5^{\prime}$ &$-0.36^{+0.45}_{-0.34}$&$-0.37^{+0.40}_{-0.34}$&$-0.36^{+0.19}_{-0.17}$&$-0.34^{+0.09}_{-0.08}$&$-0.41^{+0.11}_{-0.12}$&$0.21(21)$\\
\hline
\end{tabular}
\end{table}

In the first three columns of Tab.~\ref{tab:SMpredsOld} we show the
results produced in the SM using the model-independent  approach
described in Sec.~\ref{sec:theo}. The relevant input
parameters with their uncertainties are listed in 
Table~\ref{tab:param}, while for the Wilson coefficients in the SM we use NNLL
accuracy~\cite{Bobeth:1999mk}. The central values in the two first columns
correspond to those of the parameters and the errors to the maximum spread 
produced by the scan within the range produced by their uncertainties.
In the second column we employ a different parametrization for the soft form
factors. Instead of $T_-$ and $S$, and following the approach suggested in
~\cite{Descotes-Genon:2014uoa}, we choose $\xi_\perp=m_B/(2\,E)\,V_-$ and 
$\xi_\parallel=V_0$ to study the effect on the uncertainties introduced 
by these choices. The eq.~(\ref{eq:xiq2}) and values of the 
parameters in (\ref{eq:alphas}) and (\ref{eq:ab_pc}) also hold in this case. 
In the third column we show the results assuming that the parameters
$\vec{y}$ are normally distributed and uncorrelated, and we quote the mean and the 
$1\sigma$ CL of the distributions resulting for the observables. 

From the results of the two first columns we conclude that different
identifications of the soft form factors to the QCD ones can
produce small differences in the results, which can be reconciled with reasonable
changes in the power-correction parameters $a_X$ and $b_X$. As stressed
in Sec.~\ref{sec:theoff}, this comparison provides a test of self-consistency
to any approach based on the HQ/LE limit. Therefore, it is surprising that the
opposite conclusion was drawn from the similar analysis of 
ref.~\cite{Descotes-Genon:2014uoa}. By comparing the results obtained
in the first and third columns we observe that treating the theoretical
errors as normally distributed one obtains, at 1$\sigma$, less than $1/2$
the uncertainty produced by the scan of the parameter space.

\begin{table}[h]
\centering
\caption{Input parameters employed in the calculations of this work. 
Values for the soft form factors and power-correction parameters is
found in Sec.~\ref{sec:theo}. \label{tab:param}}
\begin{tabular}{|c|c||c|c|}
\hline
$m_{b,\,\rm{(1S)}}$ &4.65(3)~\cite{Agashe:2014kda}&$\alpha_s(M_Z)$&0.1184(7)~\cite{Agashe:2014kda}\\
$m_c$ & 1.275(25)~\cite{Agashe:2014kda}&$f_{K^*}$&220(5) MeV~\cite{Jager:2012uw}\\
$m_s$ & 0.095(5)~\cite{Agashe:2014kda}&$f_{K^*\perp}$&170(20) MeV~\cite{Jager:2012uw}\\
$\lambda$ &0.22543(77)~\cite{Charles:2004jd} & $a_1$ &0.2(2)~\cite{Beneke:2001at}\\
$A$ &0.812(19)~\cite{Charles:2004jd}&$a_2$ &0.1(3)~\cite{Beneke:2001at}\\
$\overline{\rho}$&0.145(27)~\cite{Charles:2004jd}&$f_B$&0.1905(42) GeV~\cite{Aoki:2013ldr}\\
$\overline{\eta}$&0.343(15)~\cite{Charles:2004jd}&$\lambda_{B,+}^{-1}$&2.0(5) GeV$^{-1}$~\cite{Beneke:2001at}\\
\hline
\end{tabular}
\end{table}

In the fourth and fifth columns of Tab.~\ref{tab:SMpredsOld} we show the
results of independent calculations, refs~\cite{Beaujean:2012uj}
(Bayesian) and~\cite{Descotes-Genon:2014uoa}
respectively, and in the last column the LHCb data. The predictions from 
ref~\cite{Beaujean:2013soa} (nominal priors) are the result of
a Bayesian treatment assuming normal 
distributions for the nuisance parameters. The description of the power 
corrections is different to the one employed in this work. A scale factor 
is introduced at the level of the amplitudes~\cite{Egede:2008uy} that
is normally distributed around 1 and with the standard deviation at
the $0.15\%$. 
However note that with this method the corrections to the relation 
$T_\lambda(q^2)=V_\lambda(q^2)$ are omitted and, as discussed 
in Sec.~\ref{sec:phenogen}, the uncertainties in observables like 
$P_5^{\prime}$ or $P_2$ could be underestimated at low $q^2$. 
On the other hand, ref~\cite{Descotes-Genon:2014uoa} (KMPW scheme) uses a 
parametrization of the power corrections to the form factors similar
to the one proposed here, and quotes the maximum spread of theoretical predictions.
However the power-correction parameters are not treated as uncertainties constrained only
by model-independent relations but they are in part fitted to  LCSR calculations.

We finish the comparison by noting that the uncertainties have increased as compared
to our previous work in ref.~\cite{Jager:2012uw}. This is largely explained by the improved treatment
of the nonfactorizable soft contributions used in this paper, and to a lesser extent, by the different
values used for the power-corrections parameters and $\xi_\perp(0)$. Also in the present analysis we scan
all parameters simultaneously, whereas in ref.~\cite{Jager:2012uw} they were separated in four groups 
and added in quadratures. 

\begin{sidewaystable}
\centering
\caption{Binned results in the SM for the branching fraction, the longitudinal
polarization fraction $F_L$ and the angular observables in the $P_i^{(\prime)}$ basis
(using the LHCb conventions~\cite{Aaij:2013iag,Aaij:2013qta}). For the electronic
mode we give predictions for the bin $[0.0020^{+0.0008}_{-0.0008},\,1.12^{+0.06}_{-0.06}]$~\cite{Aaij:2015dea}.
\label{tab:SMpredsNew}}
\begin{tabular}{|c|cc|ccccccc|}
\hline
Bin [GeV$^2$]&$Br$ $[10^{-8}]$&$F_L$&$P_1$&$P_2$&$P_3^{CP}$ $[10^{-4}]$&$P_4^{\,\prime}$&$P_5^{\,\prime}$&$P_6^{\,\prime}$&$P_8^{\,\prime}$\\
\hline
$[0.1,\,0.98]$&$8.6^{+4.5}_{-3.1}$&$0.26^{+0.21}_{-0.14}$&$0.03^{+0.06}_{-0.05}$&$-0.175^{+0.039}_{-0.041}$&$0.2^{+1.1}_{-0.8}$&$0.19^{+0.06}_{-0.08}$&
$0.56^{+0.13}_{-0.14}$&$0.04^{+0.08}_{-0.08}$&$0.00^{+0.09}_{-0.09}$\\
\hline
$[1.1,\,2]$&$3.4^{+2.9}_{-1.5}$&$0.68^{+0.17}_{-0.23}$&$0.04^{+0.11}_{-0.11}$&$-0.83^{+0.16}_{-0.09}$&$0.4^{+3.4}_{-2.3}$&$0.04^{+0.16}_{-0.18}$&
$0.35^{+0.30}_{-0.32}$&$0.06^{+0.19}_{-0.19}$&$0.01^{+0.11}_{-0.11}$\\
\hline
$[2,\,3]$&$3.4^{+3.4}_{-1.5}$&$0.78^{+0.13}_{-0.21}$&$0.01^{+0.12}_{-0.15}$&$-0.84^{+0.39}_{-0.14}$&$0.4^{+4.3}_{-2.9}$&$-0.19^{+0.23}_{-0.20}$&
$-0.10^{+0.47}_{-0.42}$&$0.06^{+0.26}_{-0.27}$&$0.02^{+0.09}_{-0.09}$\\
\hline
$[3,\,4]$&$3.6^{+3.8}_{-1.8}$&$0.77^{+0.14}_{-0.24}$&$-0.03^{+0.27}_{-0.27}$&$-0.21^{+0.50}_{-0.53}$&$0.3^{+3.4}_{-2.6}$&$-0.37^{+0.23}_{-0.16}$&
$-0.49^{+0.52}_{-0.36}$&$0.05^{+0.27}_{-0.28}$&$0.01^{+0.06}_{-0.06}$\\
\hline
$[4,\,5]$&$4.0^{+4.3}_{-2.1}$&$0.73^{+0.18}_{-0.28}$&$-0.06^{+0.34}_{-0.32}$&$0.30^{+0.35}_{-0.52}$&$0.2^{+2.3}_{-2.1}$&$-0.45^{+0.20}_{-0.12}$&
$-0.69^{+0.48}_{-0.30}$&$0.04^{+0.25}_{-0.26}$&$0.01^{+0.05}_{-0.05}$\\
\hline
$[5,\,6]$&$4.6^{+5.1}_{-2.6}$&$0.68^{+0.22}_{-0.30}$&$-0.07^{+0.39}_{-0.38}$&$0.59^{+0.23}_{-0.40}$&$0.1^{+1.7}_{-1.6}$&$-0.48^{+0.17}_{-0.10}$&
$-0.80^{+0.43}_{-0.27}$&$0.03^{+0.23}_{-0.24}$&$0.01^{+0.05}_{-0.06}$\\
\hline
$[1.1,\,6]$&$19^{+19}_{-9}$&$0.73^{+0.17}_{-0.25}$&$-0.02^{+0.23}_{-0.24}$&$-0.10^{+0.41}_{-0.39}$&$0.3^{+2.7}_{-1.9}$&$-0.30^{+0.21}_{-0.16}$&
$-0.38^{+0.46}_{-0.34}$&$0.05^{+0.24}_{-0.25}$&$0.01^{+0.06}_{-0.05}$\\
\hline
Electron &$23^{+10}_{-8}$&$0.12^{+0.14}_{-0.07}$&$0.03^{+0.05}_{-0.05}$&$-0.080^{+0.017}_{-0.016}$&$0.3^{+1.0}_{-0.7}$&$0.19^{+0.06}_{-0.07}$&
$0.52^{+0.12}_{-0.12}$&$0.04^{+0.07}_{-0.07}$&$0.00^{+0.08}_{-0.08}$\\
\hline
\end{tabular}
\end{sidewaystable}

\begin{sidewaystable}
\centering
\caption{Binned results in the SM for the branching fraction, the longitudinal
polarization fraction $F_L$ and the angular observables in the $P_i^{(\prime)}$ basis
(using the LHCb conventions~\cite{Aaij:2013iag,Aaij:2013qta}). The results are obtained using 
Montecarlos in which the nuisance parameters are distributed normally and we quote for each case 
the mean as central value and the $1\sigma$ intervals as error bars. For the electronic
mode we give predictions for the bin $[0.0020^{+0.0008}_{-0.0008},\,1.12^{+0.06}_{-0.06}]$~\cite{Aaij:2015dea}.
\label{tab:SMpredsGauss}}
\begin{tabular}{|c|cc|ccccccc|}
\hline
Bin [GeV$^2$]&$Br$ $[10^{-8}]$&$F_L$&$P_1$&$P_2$&$P_3^{CP}$ $[10^{-4}]$&$P_4^{\,\prime}$&$P_5^{\,\prime}$&$P_6^{\,\prime}$&$P_8^{\,\prime}$\\
\hline
$[0.1,\,0.98]$&$8.5^{+2.0}_{-1.8}$&$0.26^{+0.10}_{-0.09}$&$0.025^{+0.023}_{-0.022}$&$-0.177^{+0.020}_{-0.021}$&$0.02^{+0.34}_{-0.36}$&$0.181^{+0.029}_{-0.035}$&
$0.57^{+0.06}_{-0.06}$&$0.040^{+0.033}_{-0.033}$&$0.000^{+0.044}_{-0.043}$\\
\hline
$[1.1,\,2]$&$3.4^{+1.1}_{-0.9}$&$0.68^{+0.10}_{-0.12}$&$0.03^{+0.05}_{-0.05}$&$-0.84^{+0.07}_{-0.06}$&$0.3^{+0.8}_{-0.8}$&$0.03^{+0.07}_{-0.08}$&
$0.36^{+0.14}_{-0.14}$&$0.06^{+0.08}_{-0.08}$&$0.01^{+0.05}_{-0.05}$\\
\hline
$[2,\,3]$&$3.4^{+1.4}_{-1.0}$&$0.78^{+0.07}_{-0.11}$&$0.01^{+0.06}_{-0.07}$&$-0.83^{+0.16}_{-0.10}$&$0.3^{+0.9}_{-0.7}$&$-0.20^{+0.10}_{-0.10}$&
$-0.10^{+0.21}_{-0.20}$&$0.06^{+0.11}_{-0.11}$&$0.014^{+0.035}_{-0.035}$\\
\hline
$[3,\,4]$&$3.6^{+1.5}_{-1.1}$&$0.78^{+0.08}_{-0.11}$&$-0.04^{+0.16}_{-0.16}$&$-0.19^{+0.24}_{-0.26}$&$0.3^{+0.9}_{-0.7}$&$-0.37^{+0.10}_{-0.08}$&
$-0.49^{+0.22}_{-0.19}$&$0.05^{+0.12}_{-0.12}$&$0.013^{+0.020}_{-0.020}$\\
\hline
$[4,\,5]$&$4.0^{+1.8}_{-1.2}$&$0.73^{+0.10}_{-0.13}$&$-0.05^{+0.22}_{-0.21}$&$0.32^{+0.18}_{-0.23}$&$0.2^{+0.8}_{-0.7}$&$-0.45^{+0.08}_{-0.07}$&
$-0.69^{+0.20}_{-0.17}$&$0.3^{+0.12}_{-0.12}$&$0.011^{+0.017}_{-0.017}$\\
\hline
$[5,\,6]$&$4.6^{+1.9}_{-1.4}$&$0.67^{+0.12}_{-0.15}$&$-0.07^{+0.24}_{-0.24}$&$0.60^{+0.12}_{-0.17}$&$0.1^{+0.6}_{-0.6}$&$-0.48^{+0.08}_{-0.06}$&
$-0.80^{+0.18}_{-0.15}$&$0.02^{+0.11}_{-0.11}$&$0.01^{+0.02}_{-0.02}$\\
\hline
$[1.1,\,6]$&$19^{+8}_{-5}$&$0.73^{+0.09}_{-0.12}$&$-0.03^{+0.14}_{-0.14}$&$-0.08^{+0.19}_{-0.20}$&$0.2^{+0.6}_{-0.5}$&$-0.31^{+0.09}_{-0.08}$&
$-0.38^{+0.20}_{-0.17}$&$0.04^{+0.10}_{-0.11}$&$0.01^{+0.02}_{-0.02}$\\
\hline
Electron &$23^{+6}_{-5}$&$0.13^{+0.06}_{-0.05}$&$0.033^{+0.021}_{-0.019}$&$-0.080^{+0.008}_{-0.008}$&$0.27^{+0.32}_{-0.36}$&$0.194^{+0.026}_{-0.032}$&
$0.52^{+0.05}_{-0.05}$&$0.035^{+0.027}_{-0.027}$&$0.001^{+0.042}_{-0.042}$\\
\hline
\end{tabular}
\end{sidewaystable}

\subsection{Predictions for $B \to K^* \mu^+ \mu^-$ and $B \to K^*
  e^+ e^-$ angular observables}
In Tab.~\ref{tab:SMpredsNew}, we update our predictions
in the SM and the muonic mode for the branching fraction, the longitudinal
polarization fraction $F_L$ and the angular observables in
the $P_i^{(\prime)}$ basis (defined using the LHCb conventions~\cite{Aaij:2013iag,Aaij:2013qta}),
in the narrow binning scheme. We also
show predictions for the electronic mode in an ``effective'' low-$q^2$ 
bin~\cite{Aaij:2015dea} $[0.0020^{+0.0008}_{-0.0008},\,1.12^{+0.06}_{-0.06}]$. 
All these predictions are obtained scanning the nuisance 
parameter space and extracting the maximum spread for the errors. The
errors propagated from the uncertainties in the range of the bin in the electronic 
case are very small and neglected.

In Tab.~\ref{tab:SMpredsGauss} we show instead the predictions we
obtained using Gaussian distributions for the nuisance 
parameters, quoting the mean and the $1\sigma$ interval for the observables.

\subsection{The $B\to K^*\mu^+\mu^-$ anomaly}
\label{sec:anomaly}

As it has been stressed above, the angular observables in
$B\to K^*\ell^+\ell^-$ at low $q^2$ are usually afflicted by the 
leading power corrections to the HQ/LE factorization formula
for the decay amplitude. In order to illustrate the impact that these
can have in the phenomenological analyses, we investigate the significance
of the recently claimed tensions with the SM in $B\to K^*\mu^+\mu^-$ at low $q^2$
and employing the model-independent approach to the nonperturbative uncertainties 
presented here and in ref.~\cite{Jager:2012uw}. As discussed in 
sec.~\ref{sec:theononfact}, the bin showing the largest discrepancy,
$[4.3,\,8.63]$ GeV$^2$~\cite{Aaij:2013qta},
suffers from uncertain charm-loop contributions for
which no model-independent framework exists even in the heavy-quark
limit, as the bin extends outside the range of validity of QCD
factorization, employed both here and in~\cite{Descotes-Genon:2014uoa}. 
Therefore, we restrict
our investigation to the $CP$-averaged angular observables in the
$P_i^{(\prime)}$ basis, augmented by the $CP$ asymmetry $P_3^{
    CP}$, measured in the bin $[1,\,6]$ GeV$^2$~\cite{Aaij:2013iag,Aaij:2013qta}.

\begin{figure}[h]
\begin{tabular}{cc}
  \includegraphics[width=60mm]{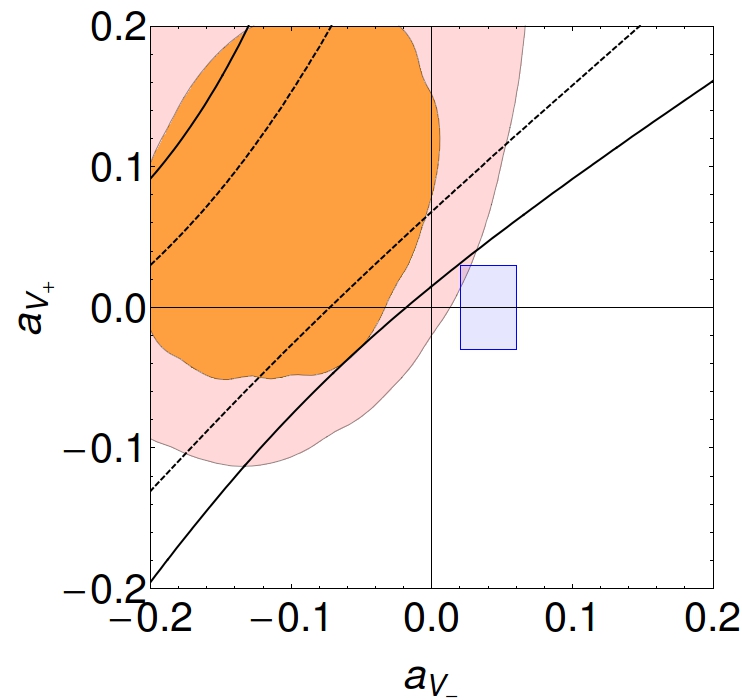}\hspace{0.8cm} & \hspace{0.8cm}  \includegraphics[width=75mm]{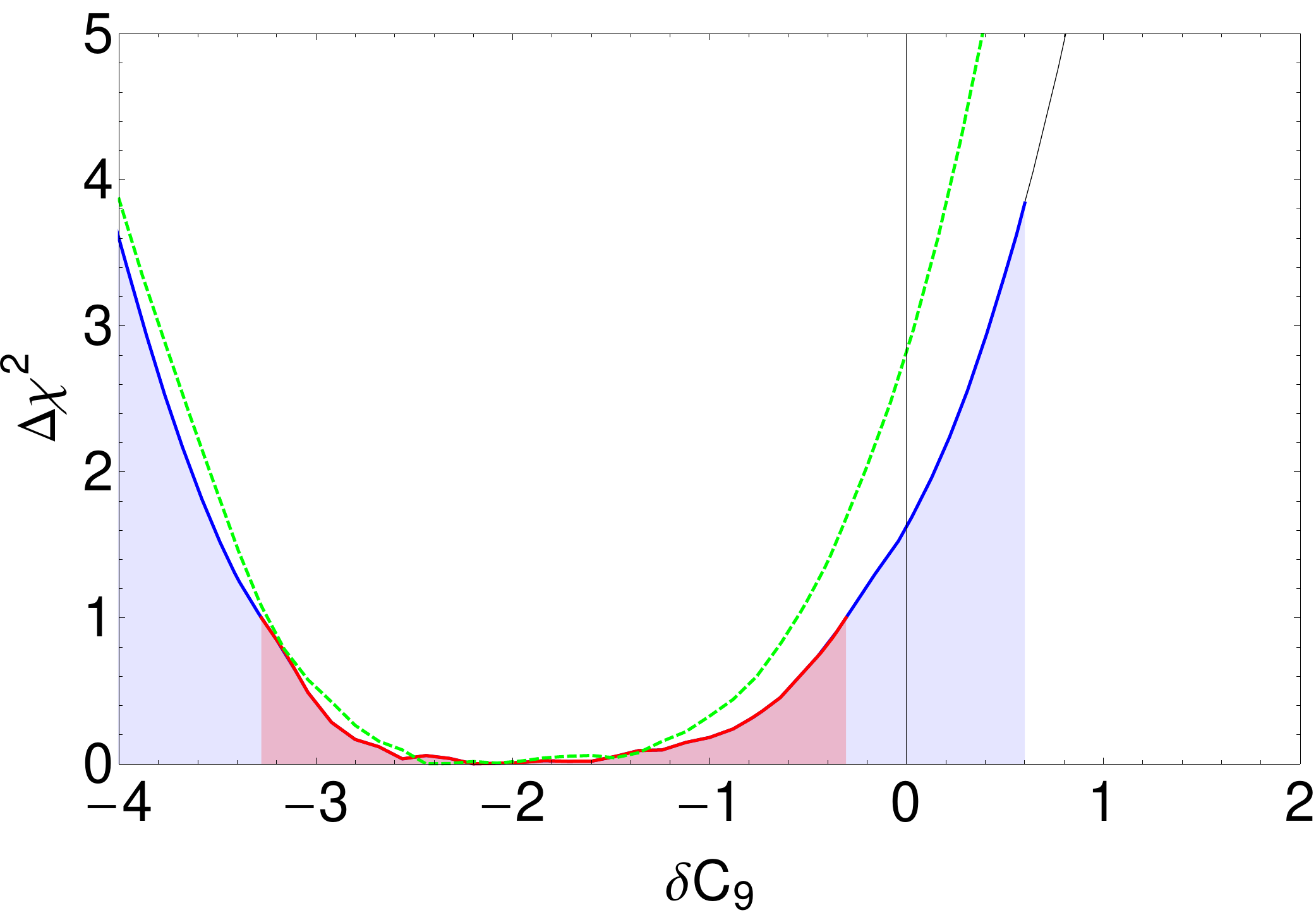} 
\end{tabular}
\caption{Graphs for the $B\to K^*\ell^+\ell^-$ anomaly. \textit{Left panel:} 
$68\%$ and $95\%$ CL bounds in the parameter space of the power corrections $a_{V_\pm}$
and for a fit in the SM. We use a profile $\chi^2$ including only the $P_i^{(\prime)}$ 
observables in the bin $[1,\,6]$ GeV$^2$. The origin of the axis corresponds to QCDF and
the small dashed box corresponds to the subspace for the LCSR of ref.~\cite{Ball:2004rg}
when the errors of $V$ and $A_1$ are combined linearly. \textit{Right panel:} Profile 
$\chi^2$ including only the angular observables in the $P_i^{(\prime)}$ basis in 
the bin $[1,\,6]$ GeV$^2$ as a function of a BSM contribution to $C_9$ and 
setting all the other Wilson coefficients to their SM values. The red and blue shades indicate
the limits for the $68\%$ and $95\%$ CL. The dashed green line corresponds to
the case in which $V_-$ and $V_0$ are used to fix the soft form factors. 
In both cases $\chi^2_{\rm min}~\sim1$.
\label{fig:C9anomaly}}
\end{figure}

On the left-hand side of Fig.~\ref{fig:C9anomaly} we show
the contours for the $\chi^2$ constructed with this angular data 
in the SM (all the Wilson coefficients set to their
SM values), as a function of the power corrections to the vector
form factors $a_{V_{\pm}}$ and where we have profiled over the
rest of the QCD parameters. The $\chi^2$ receives an important
contribution from the measured $P_5^\prime$, which in our plot 
is represented by the overlaid diagonal contours obtained
setting all the other QCD parameters to their central values. 
This is consistent with the conclusions of the different analyses 
(eg ref.~\cite{Descotes-Genon:2013wba}), and we also agree that the data 
favours a negative NP contribution to $C_9$. However, in our case the
significance is smaller, below $2\sigma$, as in our approach
the data can be accommodated quite well by reasonable values of
the power corrections. This is shown
on the right hand side of Fig.~\ref{fig:C9anomaly} where we plot 
the $\Delta\chi^2$ as a function of the contribution of NP to $C_9$
($\chi^2_{\rm min}~\sim1$) and where, $(i)$ we profile 
over all QCD parameters; $(ii)$
we set the NP contributions to all other Wilson coefficients
to zero. One can explore the dependence of these results on
some of our choices. In particular, we show also the $\chi^2$ for
the case in which we use ($V_-$, $V_0$) to fix the soft form factors.
In this case, represented by the green dashed line, and where we have 
used the same estimates for the power corrections as in our standard
choice, the agreement worsens somewhat, increasing the
significance to close to $2\sigma$. We emphasize, in line with our
above demonstration of form-factor-scheme-independence of the
observables to linear order, that this is increase in
significance is not to be interpreted as favouring this form-factor
scheme. Rather the difference in significances between the two
form-factor schemes can be taken as a measure of the uncertainty
due to higher-order power corrections.

The advantage of our parametrization of the power
corrections is that they are related to specific QCD matrix 
elements and the outcome of our analysis can be
compared directly to the results of nonperturbative calculations.
For instance, our SM fit on the left-hand side of Fig.~\ref{fig:C9anomaly}
favours a value of $a_{V_{-}}$, more generally of
  $a_{V_-} - a_{T_-}$ or equivalently a correction to the
ratio $V^\infty_-(0)/T^\infty_-(0)$ computed in QCDF,
that is negative. However, as it was advanced in
Sec.~\ref{sec:theoff}, this is a scenario that is not 
compatible with the LCSR calculation of ref.~\cite{Ball:2004rg},
where the correction is obtained with the opposite sign.
This is illustrated by the blue box in the plot, which describes
the size of the power corrections predicted by the LCSR and 
estimated as described in Sec.~\ref{sec:theoff}. 

In conclusion, the interpretation $B\to K^*\mu^+\mu^-$ anomaly 
is blurred by the sensitivity of the relevant observables to power corrections.
From the discussion above it is clear that implementing the QCD form factors in
the LCSR enhances the signal and that a careful assessment of the accuracy of the
various nonperturbative approaches seems to be still necessary to unambiguously
attribute this anomaly to NP
(for some recent developments within LCSR see~\cite{Hambrock:2013zya}).
 
\subsection{Constraints on right-handed currents}
\label{sec:C7p}

In Sec.~\ref{sec:theo} and \ref{sec:phenogen}, we argued that 
the angular observables $P_1$ and $P_3^{CP}$ stand out among the others
because they are sensitive to $H_V^+$. This makes
them zero in the SM, up to subleading power corrections or up to leading ones
that are further suppressed by a factor $q^2/m_B^2$. Thus, around
the low $q^2$ endpoint, $P_1$ and $P_3^{CP}$ are null tests of the SM 
in very good approximation, becoming very sensitive to right-handed currents
BSM entering through the electromagnetic penguin operator $Q_7^{\prime}$.

\begin{table}[h]
\caption{Error budget for the $P_1$ and $P_3^{CP}$ for the electronic mode in 
the bin $[0.002,\,1.12]$ GeV$^2$. \label{tab:elbudget}}
\begin{center}
\begin{tabular}{|c|c|ccc|}
\hline
&Result&QCDF&Fact. p.c.'s&Non-fact. p.c.'s\\
\hline
$P_1$&0.032$^{+0.055}_{-0.048}$&$^{+0.010}_{-0.003}$&$\pm0.012$&$^{+0.031}_{-0.029}$\\
$P_3^{CP}$ $[10^{-4}]$&$0.3^{+1.0}_{-0.7}$&$^{+0.5}_{-0.2}$&$^{+0.1}_{-0.2}$&$^{+0.6}_{-0.4}$\\
\hline
\end{tabular}
\end{center}
\end{table}

On the experimental side, there are the measurements of the muonic mode in
the lowest-$q^2$ bin and those in the electronic channel, which are best
due to its lower endpoint. Indeed, most of the difference between their
branching fractions in the lowest bin shown in Tab.~\ref{tab:SMpredsNew}
(roughly a factor 3) stems from events in the region between the two endpoints.
This region is especially sensitive to the physics of the photon pole and it
is where the $q^2/m_B^2$-suppression of the leading power corrections is
maximally effective. We show in Tab.~\ref{tab:elbudget} the error budget
for the SM predictions of $P_1$ and $P_3^{CP}$ in the electronic mode. As 
expected, the contribution to the uncertainty from the power corrections
to the form factors is very small and the final theoretical error is dominated
by the $(\Lambda/m_B)^2$-suppressed long-distance charm, $r_+^c$. 

The radiative decays are also obvious probes of the structure 
of the electromagnetic operators and strategies to determine the 
helicity of the photon (and therefore $C_7^{\prime}$) with these decays 
have been intensively investigated~\cite{Atwood:1997zr,
Gronau:2001ng,Gronau:2002rz,Grinstein:2004uu,Ball:2006cva,Ball:2006eu,
Kou:2010kn,Becirevic:2012dx}.
The inclusive $b \to s \gamma$ decay ratio, while theoretically
  clean, depends only quadratically on $C_7'$. However,
interference between helicity amplitudes (and linear dependence on $C_7^{\prime}$)
can  be induced by $B^0$-$\bar{B}^0$ mixing and the time dependent $CP$-asymmetry
in $B\to K^*\gamma$ and has been suggested as a clean null test of the 
SM~\cite{Atwood:1997zr}, via its sine coefficient given by:
\begin{equation}
S_{K^*\gamma}=\frac{2\,{\rm Im}\left[e^{-2\,i\,\beta}\left(H_-^*\,\bar{H}_-+
H_+^*\,\bar{H}_+\right)\right]}{|H_+|^2+|H_-|^2+|\bar{H}_+|^2+|\bar{H}_-|^2},\label{eq:SKsgamma}
\end{equation}
where  $H_\lambda$ ($\bar{H}_\lambda$) corresponds to the 
helicity amplitudes of $B\to K^*\gamma$ ($\bar{B}\to \bar{K}^*\gamma$)
and $\beta$ is the angle of the CKM unitarity triangle. These helicity
amplitudes are given in terms of the residues at $q^2=0$ of those of the semileptonic decay,
$H_V(\lambda)$, cf. eq.~(\ref{eq:radiative}), and inherit the
multiple suppression of $H_V^+$ in the SM. Applying the approach described 
in this paper for the theoretical description of the amplitudes one obtains,
in the SM:
\begin{equation}
S_{K^*\gamma}=-0.02^{+0.016}_{-0.023}, \label{eq:SKsgammaSM} 
\end{equation}
which is to be compared with the average of experimental 
measurements,
$S_{K^*\gamma}=-0.16\pm0.22$~\cite{Aubert:2005bu,Ushiroda:2006fi,Amhis:2012bh}.

\begin{figure}[h]
\begin{tabular}{cc}
  \includegraphics[width=65mm]{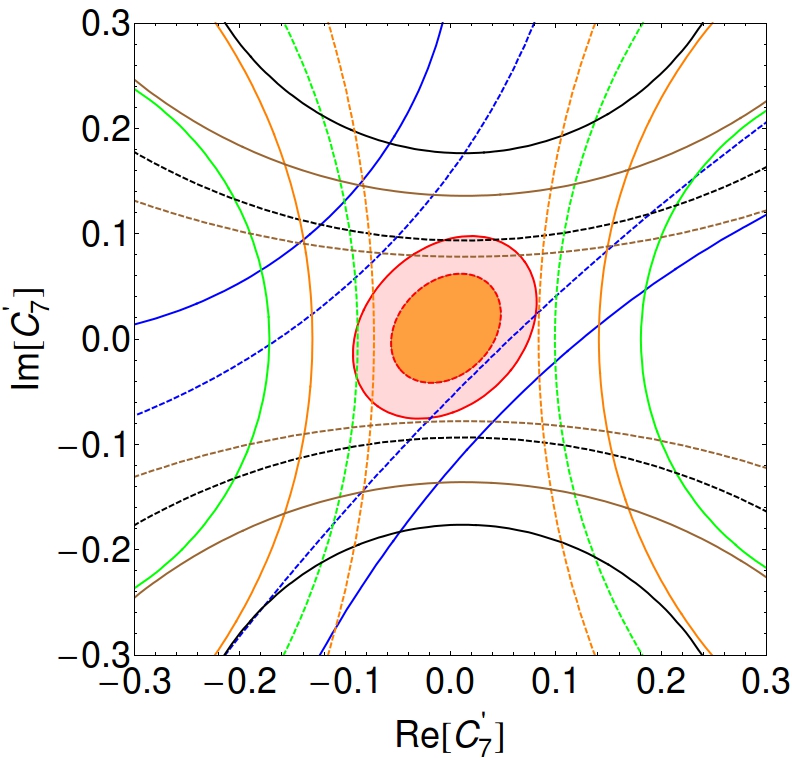}\hspace{0.8cm} & \hspace{0.8cm}  \includegraphics[width=65mm]{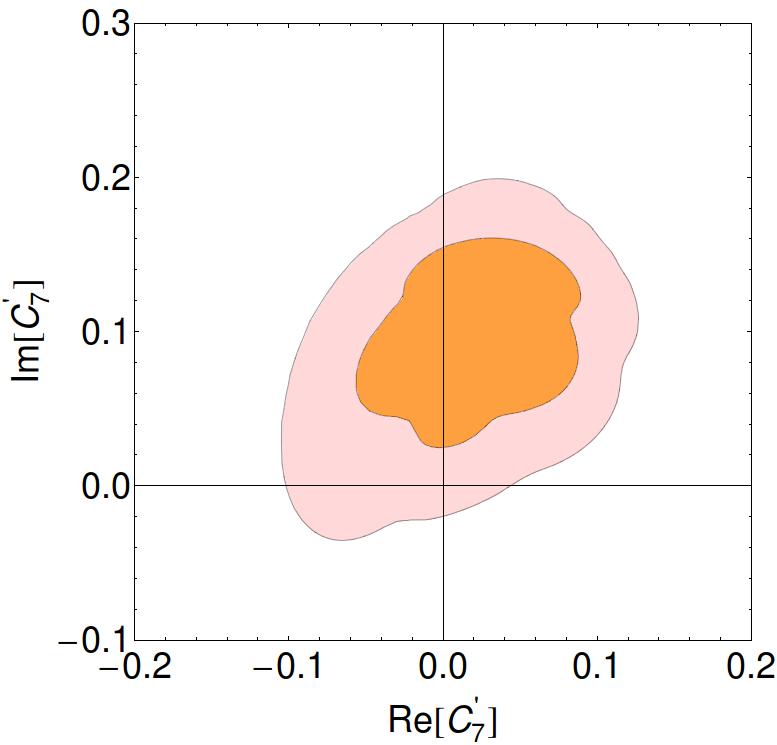} 
\end{tabular}
\caption{Bounds in the $C_7^\prime$ plane. \textit{Left panel:} 
Ideal $68\%$ and $95\%$ contour plots for the central values of the
theoretical parameters. The diagonal band corresponds to the $S_{K^*\gamma}$
measurement and the vertical and horizontal ones to hypothetical null measurements
of $P_1$ and $P_3^{CP}$, respectively, with an assumed experimental precision of $0.1$.
The green and black lines are for the muonic mode and the brown and orange 
for the electronic one.
\textit{Right panel:} Current bounds at $68\%$ and $95\%$ CL in the
$C_7^\prime$ plane using all the current data of $B\to K^*\mu^+\mu^-$ in the lower 
bins $[0.1,\,2]$ and $[2,\,4.3]$ GeV$^2$ and of $B\to K^*\gamma$ and $B\to X_s\gamma$.
We use the profile likelihood method and set all other Wilson coefficients
to their SM values.
\label{fig:C7p}}
\end{figure}

In Fig.~\ref{fig:C7p} we display contour plots in the $C_7^\prime$ plane. 
On the left-hand side we show the ideal bounds obtained for the central
values of the theoretical parameters and assuming an experimental precision
of $0.25$ for $P_1$ and $P_3^{CP}$ in both the electronic and muonic channels.
We also show the constraint obtained from the current measurements of $S_{K^*\gamma}$.
On the right hand side we show the current bound that is obtained using all
the available experimental data of $B\to K^*\ell^+\ell^-$ in the lower bins 
$[0.1,\,2]$ and $[2,\,4.3]$ GeV$^2$ and of $B\to K^*\gamma$. We also use the 
branching fraction of $B\to X_s\gamma$, $\mathcal{B}(B\to X_s\gamma)$ that,
depending quadratically on $C_7^\prime$, becomes the dominant contribution
to the $\chi^2$ for large values of the coefficient. We profile over all the
QCD parameters, and we set all the other Wilson coefficients to their SM values. 

We conclude that $P_1$ and $P_3^{CP}$ conform, in combination
with $S_{K^*\gamma}$ and $\mathcal{B}(B\to X_s\gamma)$, and neglecting NP 
contributions to the phase of the $B_d$ mixing amplitude, a basis of clean 
observables that completely determine $C_7$ and $C_7^{\prime}$ from
experiment, with the simple expressions given in
\cite{Kou:2010kn,Becirevic:2012dx} being protected from QCD
uncertainties to a high degree.

With the small theoretical uncertainties in the SM predictions, one expects that
the determination of these Wilson coefficients will be dominated by the experimental
errors. 
In this regard, and as shown in the left-hand panel of Fig.~\ref{fig:C7p}, the 
measurements provided by the electronic mode are very promising. It is also worth 
pointing out in the right-hand panel of Fig.~\ref{fig:C7p} the small discrepancy 
with the SM in the imaginary part that is driven by the current measurement of 
the angular observable $A_9$ in the muonic mode.
\footnote{Since this discussion is meant to be an 
illustration of the impact of the approach of this paper in the phenomenology,
we obtained our experimental $P_3^{CP}$ from the measured $A_9$ and $F_L$ via 
$P_3^{CP}=A_9/(1-F_L)$, propagating errors quadratically and ignoring experimental
correlations.} 

\begin{figure}[h]
\begin{tabular}{cc}
  \includegraphics[width=75mm]{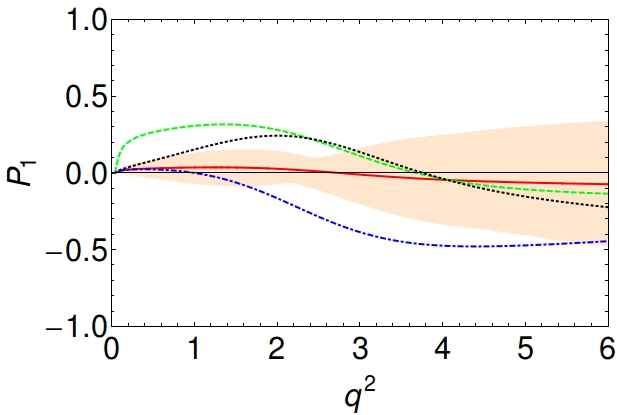}\hspace{0.5cm} & \hspace{0.5cm}  \includegraphics[width=75mm]{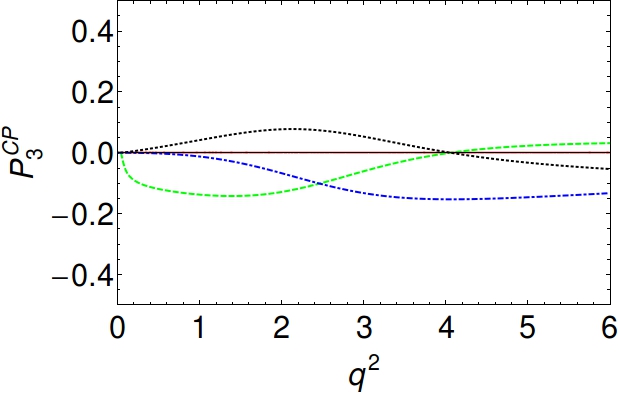} 
\end{tabular}
\caption{Sensitivity of $P_1(q^2)$ and $P_3^{CP}(q^2)$ to right-handed quark currents. 
The SM predictions are the solid (red) lines and the theoretical uncertainty,
represented by the band, is obtained taking the maximum spread of theoretical 
predictions. In the plot of $P_1(q^2)$ the NP scenarios correspond to 
$C_7^\prime=-0.05$ (dashed green), $C_9^\prime=-1$ (dotted black) and 
$C_{10}^\prime=1$ (dot-dashed blue). In the plot of $P_3^{CP}(q^2)$, we
show $C_7^\prime=-0.05$ (dashed green), $C_9^\prime=-1$ (dotted black) and 
$C_{10}^\prime=1$ (dot-dashed blue). In the plot of $P_3^{CP}(q^2)$, we 
plot $C_7^\prime=0.05\,i$ (dashed green), $C_9^\prime=-e^{i\frac{\pi}{4}}$ (dotted black) and 
$C_{10}^\prime=e^{i\frac{\pi}{4}}$ (dot-dashed blue).
\label{fig:P1P3CPsens}}
\end{figure}

At higher $q^2$, $P_1$ and $P_3^{CP}$ are also affected 
by $\mathcal{O}(\Lambda/m_B)$ power corrections induced, in this case, chiefly  
by the vector form factor $V_+(q^2)$. However, given their specific sensitivity to right-handed 
quark currents they could also serve to probe $C_{9}^\prime$ and $C_{10}^\prime$, 
especially if these are as large as recently discussed in the literature~\cite{Altmannshofer:2014rta}.
In order to demonstrate this we show in Fig.~\ref{fig:P1P3CPsens} the $q^2$ dependence 
of $P_1$ and $P_3^{CP}$ in the SM and compared to the results in different scenarios of
NP involving corrections to $C_{9}^\prime$ and $C_{10}^\prime$ of order 1.
As discussed in Sec.~\ref{sec:stats}, the correlations with other observables
in terms of the same relevant QCD parameters (cf. $a_{V_+}$) will increase the
sensitivity  of $P_1$ and $P_3^{CP}$ to $C_{9}^\prime$ and $C_{10}^\prime$ in
the global fits.

\section{Lepton-universality ratios}
\label{sec:LURs}

In light of the recent hints on lepton universality violation (LUV)
in various $b\to s\ell\ell$ measurements it
becomes important to investigate the opportunities that $B\to K^*\ell\ell$ offers
for the confirmation and characterization of the effect. In fact, one can introduce
lepton universality ratios involving the muonic and the electronic modes for a 
given observable. These are extremely clean since up to kinematical $\mathcal{O}(m_\mu^2/q^2)$
corrections, they are equal to 1 in the SM and in lepton-universal NP scenarios.
Let us start defining the ratio of 
the rates~\cite{Hiller:2003js}:
\begin{equation}
R_{K^*}=\frac{\mathcal{B}(B\to K^*\mu^+\,\mu^-)}{\mathcal{B}(B\to K^*e^+\,e^-)}. \label{eq:RKs}
\end{equation}
Next, we can use the angular distribution of the final $K$ and $\pi$ produced in 
the $K^*$ decay to separate the total $B\to K^*\ell\ell$ rate into the contributions of
the decays into transversely or longitudinal $K^*$ mesons. The balance between these two
in the total decay rate is often measured by the polarization fractions $F_L=1-F_T$~\cite{Aaij:2013iag}.
Here we prefer to use the integrated ``longitudinal'' and ``transversal'' rates, 
$d\,\Gamma/dq^2\,F_{L,T}$, and construct the corresponding lepton-universality ratios:
\begin{eqnarray}
R_{K^*_X}=\frac{\mathcal{B}(B\to K^*_X\mu^+\,\mu^-)}{\mathcal{B}(B\to K^*_Xe^+\,e^-)}.\hspace{1cm} X=L,\,T.
\label{eq:RKsX}
\end{eqnarray}
Finally, one can define the ratio of the different angular observables that we
define as:
\begin{equation}
R_i=\frac{\langle \Sigma_i^\mu\rangle}{\langle\Sigma_i^e\rangle},\label{eq:Ri} 
\end{equation}
where $\Sigma^{\ell}_i$ stands for the given observable with the leptons $\ell$
in the notation for the $CP$ averages of~\cite{Jager:2012uw} and with 
the brackets indicating that the angular observables have been integrated 
over certain $q^2$ region. In the discussion below, we will
use the same labels for the $q^2$-dependent observables obtained 
replacing the integrated rates in eqs.~(\ref{eq:RKs}), (\ref{eq:RKsX}) and
(\ref{eq:Ri}) by the corresponding differential ones.

As it has been concluded in various analyses~\cite{Alonso:2014csa,Hiller:2014yaa},
the LUV signal ought to be produced by lepton-dependent semileptonic
operators $Q^{(\prime)}_{9,10}$. In the following, we will discuss
the lepton-universality ratios in scenarios assuming that the electronic mode
is SM-like and with the NP affecting only the muonic operators. This is supported
by the current (rather unprecise) electronic data set and it would 
also fit a possible NP contribution to the $B\to K^*\mu^+\mu^-$ anomaly
discussed in sec.~\ref{sec:anomaly}~\cite{Alonso:2014csa,Hiller:2014yaa,
Ghosh:2014awa,Hurth:2014vma,Hiller:2014ula,Gripaios:2014tna}. Nevertheless, note that
these ratios are only sensitive to the differences of the Wilson coefficients for the two
leptons. We will study three scenarios: $A$ in which $\delta C_9^\mu=-1$; $B$ where 
$\delta C_{10}^\mu=-1$; and $C$ an $SU(2)_L$-doublet scenario with 
$\delta C_9^\mu=-\delta C_{10}^\mu=-0.5$. As discussed in~\cite{Hiller:2014yaa}, 
these are all allowed by the LUV measurement in $B^+\to K^+\ell\ell$. In 
this sense, a positive and significant NP contribution to $C_{10}$ does not
seem to be ruled out in some global analyses especially if it comes in a $SU(2)_L$
combination with $C_{9}$~\cite{Altmannshofer:2014rta}. In order to study
plausible solutions based on quark right-handed currents, we will also consider,
in one particular lepton-universality ratio, the ``primed''
scenarios where $\delta C_{i}^\mu\to \delta C_{i}^{\mu\;\prime}$.

\begin{figure}[h]
\begin{tabular}{cc}
  \includegraphics[width=75mm]{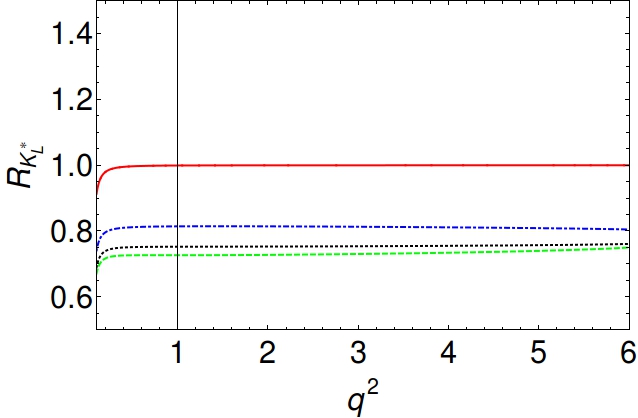}\hspace{0.5cm} & \hspace{0.5cm}  \includegraphics[width=75mm]{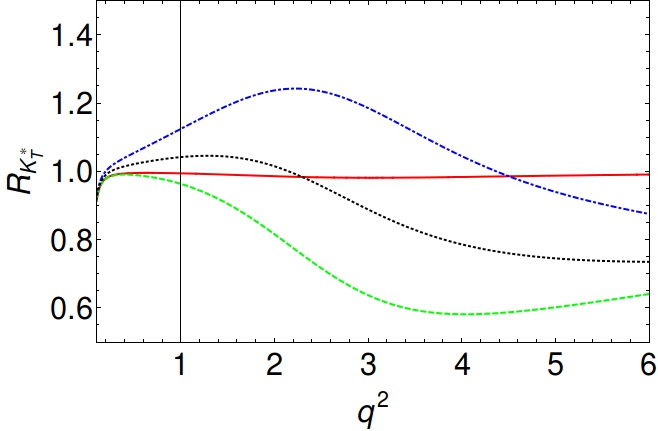}\\ 
  \includegraphics[width=75mm]{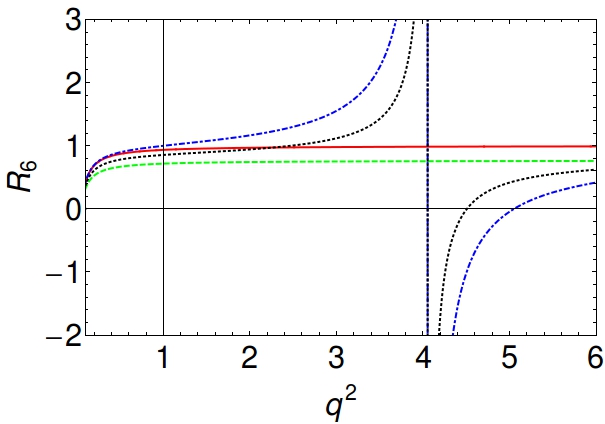}\hspace{0.5cm} & \hspace{0.5cm}  \includegraphics[width=75mm]{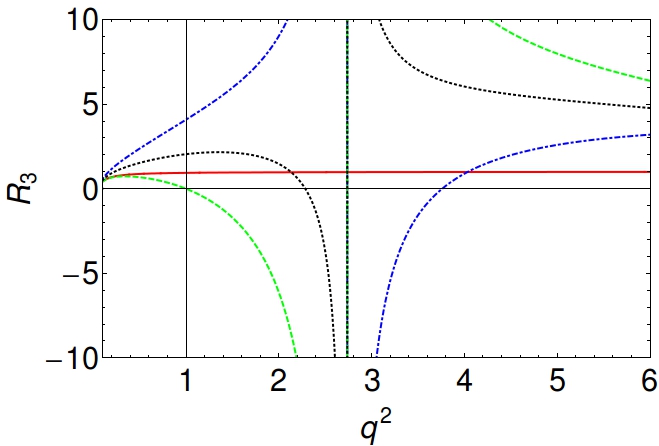}
  \end{tabular}
\caption{Lepton-universality ratios for $B^{0}\to K^{*0}\ell\ell$
in the SM and in different NP scenarios. 
The SM is the solid (red), $A$ the dashed (blue), $B$ the dot-dashed (green)
and $C$ the dotted (black) lines. In the right-hand plot of the lower panel 
($R_3$) we show, instead, the $A^\prime$, $B^\prime$ and $C^\prime$ scenarios.
\label{fig:LURs}}
\end{figure}

In Fig.~\ref{fig:LURs} we show a selection of the ratios~(\ref{eq:RKsX})
and~(\ref{eq:Ri}) plotted in terms of the $q^2$-dependent differential
rates. The first plot, on the left-hand side of the upper panel, shows the
effects of the different NP scenarios
on the longitudinal rate. This is analogous to the one in 
$B\to K\ell^+\ell^-$, and therefore a signal equivalent to the one measured in 
$R_K$ should be also found in $R_{K^*_L}$. On the other hand, the effect in the
ratio of the transversal rate is intrinsically different 
to the one probed by $R_K$. Indeed, $R_{K^*_T}$ depends on $|H_V^-|^2$,
which includes an interference term, $C_9\,C_7/q^2$, that is negative in the 
SM. Therefore, a contribution $\delta C_9<0$ is expected to reduce the
interference, increasing the transversal rate at low $q^2$. This is what we observe
on the right-hand side of upper panel in Fig.~\ref{fig:LURs}, where, in the 
scenario $A$, $R_{K^*_T}>1$ for most of the low $q^2$ region and up to the point
where the term $|C_9^\mu|^2$ dominates and makes $R_{K^*_T}<1$. Scenario $B$
involves only a reduction of the quadratic term $|C_{10}^\mu|^2$ and
therefore it causes an overall reduction of $R_{K^*_T}$ with respect to 1. 

Besides that, the ratios $R_i$ between different coefficients of the angular
observables offer unique opportunities to investigate LUV.
Some of these coefficients, like $I_6(q^2)$ (the one entering $A_{FB}$ and
$P_2$) and $I_5(q^2)$ (entering $P_5^\prime$) have zeroes at low $q^2$ due
to the cancellations between $C_9^\ell$ and $C_7/q^2$ at work within $H_V^-$ 
(see eg~\cite{Beneke:2001at}). Therefore, a displacement of the zero-crossings
between a muonic angular coefficient and its electronic counterpart
would be also an unambiguous signal of LUV, as all long-distance QCD
effects cancel out. Defining:
\begin{equation}
   \Delta_0^i \equiv (q^2_0)_{I_i}^{(\mu)} - (q^2_0)_{I_i}^{(e)} ,
\end{equation}
an observation of a nonvanishing $\Delta^0_i$ would provide provide sensitivity to LUV
in $\delta C_{9\perp}$ (primarily for $i=6$), to $\delta C_{9\perp} +
\delta C_{9\parallel}$ (primarily for $i=5$), and in addition to $\delta
C_{10}$ ($i=4$).
 
In particular, these would manifest as singularities in the $q^2$-dependence
of the corresponding lepton-universality ratios. As an example, we show on the
left-hand side of the lower panel of Fig.~\ref{fig:LURs}, $R_6(q^2)$ in
the SM compared to the different NP scenarios. As anticipated, in $A$ and $C$,
one finds singularities with the negative contributions to $C_9^\mu$ shifting
the zeroes of the muonic mode to higher $q^2$. In scenario $B$, involving only
a shift to $C_{10}^\mu$, the positions of the zeroes do not change
and the effect chiefly consists of a sustained 
(in $q^2$) reduction of $R_6$ with respect to 1.
Thus, one expects
that ratios $R_6$ taken from rates integrated up to $q^2\simeq4$ GeV$^2$
will be very sensitive to LUV entering through $C_9^\ell$. A very similar
plot and discussion can be made for $R_5$, with the exception that the zeroes
of $I_5(q^2)$ are at rather lower $q^2$, and for $R_4$, whose zero
is also sensitive to LUV in $C_{10}^\ell$. 
  
\begin{table}[t]
\caption{Binned predictions for the lepton-universality ratios in $B^{0}\to K^{*0}\ell\ell$.
Uncertainties in the SM are at the permille level.\label{tab:LURs}}
\begin{center}
 \begin{tabular}{|c|c|c|ccc|}
 \hline
 ~~Ratio~~ &~~Bin [GeV$^2$]~~&~~~~~SM~~~~~&~~~~$A$~~~~&~~~~$B$~~~~&~~~~$C$~~~~\\
 \hline
 &[0.1, 1]&0.981&0.98&0.92&0.95\\
 \multicolumn{1}{|c|}{\raisebox{2.5ex}[1.5pt]{$R_{K^*}$}}&[1, 4]&0.996&0.93&0.75&0.82\\
 \hline
 &[0.1, 1]&0.991&0.81&0.72&0.75\\
 \multicolumn{1}{|c|}{\raisebox{2.5ex}[1.5pt]{$R_{K^*_L}$}}&[1, 4]&0.999&0.81&0.73&0.75\\
 \hline
 &[0.1, 1]&0.979&1.02&0.97&1.00\\
 \multicolumn{1}{|c|}{\raisebox{2.5ex}[1.5pt]{$R_{K^*_T}$}}&[1, 4]&0.987&1.18&0.79&0.97\\
 \hline
 $R_6$&[1, 4]&0.975&1.31&0.75&1.01\\
 \hline
 \end{tabular}
 \end{center}
 \end{table}

On the right-hand side of the lower panel of Fig.~\ref{fig:LURs}, we show 
a lepton-universality ratio, $R_3$, that is qualitatively different to the 
previous ones. The $R_3$ involves the same angular observable as $P_1$, 
with its strong sensitivity to right-handed currents, but with an almost
exact cancellation of hadronic uncertainties in the ratio. In fact, this is 
shown in the plot, where the sensitivity is enhanced even more by the smallness
of $I_3^\ell(q^2)$ in the SM and the fact that it has a zero which depends
on $C_9^\prime$ and $C_{10}^\prime$. This is turn also means that
obtaining a useful measurement of $R_3$ could prove challenging as it
requires non-null measurements of 
$I_3^\ell(q^2)$, which would already indicate NP (cf. sec.~\ref{sec:C7p})
or require very high statistics.

In Tab.~\ref{tab:LURs} we present various binned predictions in the SM and
in the different NP scenarios discussed above. All of the latter produce a 
generalized reduction of $R_{K^*_L}$ by a $\sim25\%$, which is equivalent 
to the one experienced by $R_K$, although now the characteristic 
dependence of $R_{K^*_T}$ on $C_9^\mu$ allows to distinguish among them. 
In $A$ we would observe a increment of $R_{K^*_T}$, in $B$ a $~20\%$ decrease 
and, in $C$, a value that is very similar to the SM. In fact we have chosen
the bin [1, 4] GeV$^2$ because it maximizes this sensitivity of $R_{K^*_T}$
to interference of $C_9^\mu$ with the photon pole. In anticipation
to the incoming LHCb measurements of the angular observables in both, the muonic and
electronic channels, we also show the results in the [0.1, 1] GeV$^2$ bin.
In this case, $R_{K^*_T}$ does not seem to be very sensitive to the NP scenarios
studied, although the reduction in $R_{K^*_L}$ is similar to the one in the larger
bin.

Finally, we also give results for $R_6$ in [1, 4] GeV$^2$ to demonstrate
the interest that measurements of the $R_i$ ratios could have in the future. 
Nevertheless, we stress that the real potential of these observables lies in the
measurement of their $q^2$ dependence. In fact, the predictions in the SM in
large bins can have an \textit{infinite} uncertainty because, for certain values of
the nuisance parameters, the zeroes in $I_i$ are such that $\langle I_i^e\rangle=0$.
This problem, of course, disappears with the measurement of $R_i$ in a sufficiently
dense array of bins.

\section{Conclusions}
\label{sec:Concs}

We have revisited a model-independent approach to the nonperturbative
uncertainties in $B\to K^* \ell^+\ell^-$ at large recoil~\cite{Jager:2012uw} and
in the light of the new data and recent anomalies. Our approach places the
heavy-quark/large-energy limit of QCD, with its well defined predictions and
simplifications, in its core. The power corrections are parametrised incorporating
only exact (model-independent) constraints in QCD. Beyond that, we treat the 
resulting parameters as flat errors in the amplitudes whose ranges we estimate by 
power-counting arguments, in case of the form factors or using explicit model
calculations, for the nonfactorizable terms. 

With this approach to the QCD uncertainties we intend to minimize
the amount of information needed on the value and correlations of the
hadronic matrix elements that should be, otherwise, obtained from 
nonperturbative frameworks like the light-cone sum rules. The reason
for this is two-fold: \textit{(i)} We can classify angular observables
according to their sensitivity to specific NP and hadronic effects, 
especially the power corrections. \textit{(ii)} We make manifest the 
fact that the experimental data will not only test the short-distance structure
of the SM but also, simultaneously, our understanding of the hadronic matrix 
elements involved. Therefore, in the event of a disagreement with the data and before
claiming discovery of NP, it will be of utmost importance to find schemes to dissect,
and hopefully, rule out the not-fully-understood nonperturbative dynamics as
the culprit. 

We have then updated our predictions for the muonic and electronic channels
and discussed various classes of observables according to 
their theoretical cleanness and specific sensitivities to NP. 
First, we demonstrated that the observables
in the $P_i^{(\prime)}$ basis generally suffer from leading power corrections 
and argued that their theoretical uncertainties need to be carefully assessed.
These conclusions extend also to the $S_i$ basis in which the 
hadronic uncertainties generally do not cancel in the heavy-quark limit.
We then ratified the findings of previous studies about the importance
of $P_5^\prime$ (or $S_5$) in the angular analysis performed with 1 fb$^{-1}$ of LHCb data.
Our fits also favour negative NP contributions to $C_9$ although, contrary to
the conclusions of other analyses (e.g. refs.~\cite{Descotes-Genon:2013wba,Altmannshofer:2014rta}),
we find a pull of the angular obsevables of $B\to K^*\mu\mu$ at low $q^2$
(below 6 GeV$^2$) with respect to the SM of only about $2\sigma$. The better consistency
in our approach is achieved with an allowed region of the nuisance (QCD) parameter
space that is selected by the $R$fit algorithm. Similar conclusions are 
obtained in the bayesian analysis of ref.~\cite{Beaujean:2013soa} where the QCD parameters 
are also allowed to float in the fits. Our framework proves most useful
by pointing to the specific form of the QCD hadronic matrix elements that is needed.
In this sense, we discussed that agreement with the SM requires power corrections
to the ratio of the tensor and vector form factors, 
$T_-$ and $V_-$, that are not in good agreement with the ones 
obtained in light-cone sum rule calculations.

This discussion, in turn, emphasizes the value that two of these
observables have. Indeed, $P_1$ and $P_3^{CP}$ only receive contributions
from the leading power corrections that are further suppressed around the
low-$q^2$ endpoint, providing a theoretically clean window to right-handed
currents BSM. Furthermore, the experimental prospects for the dedicated 
measurement of this region, especially with the electronic modes, is very 
promising. We argued that in combination with the radiative decays,
these two observables provide a sufficient set of constraints to accurately 
determine the Wilson coefficients $C_7^\prime$ and $C_7$ from experiment. We
determined the bounds in the $C_7^\prime$ plane that can be obtained with
the current measurements on radiative decays and the angular observables in
the muonic channel to illustrate this. 

Finally, we proposed various lepton-universality ratios
and relative shifts in zero-point crossings using the angular
distributions of the muonic and electronic channels. These are all very  
accurately predicted in the SM and have a rich structure in terms of lepton-dependent
Wilson coefficients. Ratios based on the transversal and longitudinal 
contributions to the decay rate are very useful to distinguish among different
scenarios involving the semileptonic operators. In particular, $R_{K^*_L}$ is similar
to the ratio in the kaonic decay, $R_K$, whereas $R_{K^*_T}$ shows an interesting 
dependence on $C_9^\ell$ due to the interference with the photon pole in $H_V^-$.
Ratios involving the angular coefficient $I_{4,5,6}(q^2)$
are particularly interesting because of their zeroes that depend on $C_9$ and $C_{10}$ and LUV
produces the appearance of singularities in the rates.
We discussed binned predictions in various NP scenarios and concluded that
these lepton universality ratios have a great potential in terms of the discovery
and shaping of the presumed lepton-universality interactions beyond the SM.

\section{Acknowledgements}

We want to thank M.~Borsato, T.~Feldmann, B.~Grinstein, Y.~Grossman, G.~Hiller,
A.~Khodjamirian, J.~LeFrancois, Z.~Ligeti, J.~Matias, M.~Neubert, P.~Owen,
K.~Petridis, M.-H.~Schune, D.~Straub and R.~Zwicky for useful
discussions. S.J.\ acknowledges support from UK STFC
  under grant ST/L000504/1 and from the NExT institute.
J.M.C has received funding from the People Programme (Marie 
Curie Actions) of the European Union's Seventh Framework Programme 
(FP7/2007-2013) under REA grant agreement n PIOF-GA-2012-330458 and acknowledges
the Spanish Ministerio de Econom\'ia y Competitividad and European FEDER funds 
under the contract FIS2011-28853-C02-01 for support.

\noindent {\bf Note added.}
After this work was completed and submitted to the arXiv,
References \cite{Haba:2015gwa,Hofer:2015kka,Descotes-Genon:2015hea} appeared
concerning BSM searches with $B \to V \ell^+ \ell^-$ decays.
Furthermore, ref.~\cite{Straub:2015ica} appeared which updates the
LCSR predicions of \cite{Ball:2004rg}. Interestingly, while the tensor
form factor $T_-(0)=T_1(0)$ in \cite{Straub:2015ica} is significantly
lower than in \cite{Ball:2004rg} and now leads to a SM prediction for $BR(B \to K^* \gamma)$ in
agreement with experiment, the ratio $V_-(0)/T_-(0)$  receives no
significant change and remains
at variance with the power corrections inferred from data (if the SM is assumed).

\end{document}